\documentclass{eptcs}

\makeatletter
\@ifundefined{lhs2tex.lhs2tex.sty.read}%
  {\@namedef{lhs2tex.lhs2tex.sty.read}{}%
   \newcommand\SkipToFmtEnd{}%
   \newcommand\EndFmtInput{}%
   \long\def\SkipToFmtEnd#1\EndFmtInput{}%
  }\SkipToFmtEnd

\newcommand\ReadOnlyOnce[1]{\@ifundefined{#1}{\@namedef{#1}{}}\SkipToFmtEnd}
\usepackage{amstext}
\usepackage{amssymb}
\usepackage{stmaryrd}
\DeclareFontFamily{OT1}{cmtex}{}
\DeclareFontShape{OT1}{cmtex}{m}{n}
  {<5><6><7><8>cmtex8
   <9>cmtex9
   <10><10.95><12><14.4><17.28><20.74><24.88>cmtex10}{}
\DeclareFontShape{OT1}{cmtex}{m}{it}
  {<-> ssub * cmtt/m/it}{}

\DeclareFontShape{OT1}{cmtt}{bx}{n}
  {<5><6><7><8>cmtt8
   <9>cmbtt9
   <10><10.95><12><14.4><17.28><20.74><24.88>cmbtt10}{}
\DeclareFontShape{OT1}{cmtex}{bx}{n}
  {<-> ssub * cmtt/bx/n}{}

\newcommand{\Conid}[1]{\mathit{#1}}
\newcommand{\Varid}[1]{\mathit{#1}}
\newcommand{\anonymous}{\kern0.06em \vbox{\hrule\@width.5em}}

\newcommand{\bind}{\mathbin{>\!\!\!>\mkern-6.7mu=}}
\newcommand{\sequ}{\mathbin{>\!\!\!>}}
\renewcommand{\leq}{\leqslant}

\usepackage{polytable}

\@ifundefined{mathindent}%
  {\newdimen\mathindent\mathindent\leftmargini}%
  {}%

\def\resethooks{%
  \global\let\SaveRestoreHook\empty
  \global\let\ColumnHook\empty}
\newcommand*{\savecolumns}[1][default]%
  {\g@addto@macro\SaveRestoreHook{\savecolumns[#1]}}
\newcommand*{\restorecolumns}[1][default]%
  {\g@addto@macro\SaveRestoreHook{\restorecolumns[#1]}}
\newcommand*{\aligncolumn}[2]%
  {\g@addto@macro\ColumnHook{\column{#1}{#2}}}

\resethooks

\newcommand{\onelinecommentchars}{\quad-{}- }
\newcommand{\commentbeginchars}{\enskip\{-}
\newcommand{\commentendchars}{-\}\enskip}

\newcommand{\visiblecomments}{%
  \let\onelinecomment=\onelinecommentchars
  \let\commentbegin=\commentbeginchars
  \let\commentend=\commentendchars}

\newcommand{\invisiblecomments}{%
  \let\onelinecomment=\empty
  \let\commentbegin=\empty
  \let\commentend=\empty}

\visiblecomments

\newlength{\blanklineskip}
\newcommand{\hsindent}[1]{\quad}
\let\hspre\empty
\let\hspost\empty

\EndFmtInput
\makeatother
\ReadOnlyOnce{polycode.fmt}%
\makeatletter

\newcommand{\hsnewpar}[1]%
  {{\parskip=0pt\parindent=0pt\par\vskip #1\noindent}}

\newcommand{\hscodestyle}{}

\newcommand{\sethscode}[1]%
  {\expandafter\let\expandafter\hscode\csname #1\endcsname
   \expandafter\let\expandafter\endhscode\csname end#1\endcsname}

  {\par\noindent
   \advance\leftskip\mathindent
   \hscodestyle
   \let\\=\@normalcr
   \let\hspre\(\let\hspost\)%
   \pboxed}%
  {\endpboxed\)%
   \par\noindent
   \ignorespacesafterend}

  {\hsnewpar\abovedisplayskip
   \advance\leftskip\mathindent
   \hscodestyle
   \let\hspre\(\let\hspost\)%
   \pboxed}%
  {\endpboxed%
   \hsnewpar\belowdisplayskip
   \ignorespacesafterend}

  {\hsnewpar\abovedisplayskip
   \advance\leftskip\mathindent
   \hscodestyle
   \let\\=\@normalcr
   \(\pboxed}%
  {\endpboxed\)%
   \hsnewpar\belowdisplayskip
   \ignorespacesafterend}

\newcommand{\plainhs}{\sethscode{plainhscode}}

\plainhs

  {\hsnewpar\abovedisplayskip
   \advance\leftskip\mathindent
   \hscodestyle
   \let\\=\@normalcr
   \(\parray}%
  {\endparray\)%
   \hsnewpar\belowdisplayskip
   \ignorespacesafterend}

  {\parray}{\endparray}

  {\(\parray}{\endparray\)}

\def\codeframewidth{\arrayrulewidth}
\RequirePackage{calc}

  {\parskip=\abovedisplayskip\par\noindent
   \hscodestyle
   \arrayrulewidth=\codeframewidth
   \tabular{@{}|p{\linewidth-2\arraycolsep-2\arrayrulewidth-2pt}|@{}}%
   \hline\framedhslinecorrect\\{-1.5ex}%
   \let\endoflinesave=\\
   \let\\=\@normalcr
   \(\pboxed}%
  {\endpboxed\)%
   \framedhslinecorrect\endoflinesave{.5ex}\hline
   \endtabular
   \parskip=\belowdisplayskip\par\noindent
   \ignorespacesafterend}

\newcommand{\framedhslinecorrect}[2]%
  {#1[#2]}

  {\(\def\column##1##2{}%
   \let\>\undefined\let\<\undefined\let\\\undefined
   \newcommand\>[1][]{}\newcommand\<[1][]{}\newcommand\\[1][]{}%
   \def\fromto##1##2##3{##3}%
   }{\) }%

  {\let\orighscode=\hscode
   \let\origendhscode=\endhscode
   \def\endhscode{\def\hscode{\endgroup\def\@currenvir{hscode}\\}\begingroup}
   \orighscode\def\hscode{\endgroup\def\@currenvir{hscode}}}%
  {\origendhscode
   \global\let\hscode=\orighscode
   \global\let\endhscode=\origendhscode}%

\makeatother
\EndFmtInput

\providecommand{\event}{MSFP 2012} 
\usepackage{breakurl}             
\usepackage{amsmath}
\usepackage{stmaryrd}
\usepackage{multicol}
\usepackage{amsthm}
\usepackage{paralist}
\usepackage{semantic}
\usepackage{units}
\usepackage{xy}
\xyoption{all}

\newtheorem{definition}{Definition}

\title{Evaluation strategies for monadic computations}
\author{Tomas Petricek
\institute{Computer Laboratory\\
University of Cambridge\\
United Kingdom}
\email{tomas.petricek@cl.cam.ac.uk}
}
\def\titlerunning{Evaluation strategies for monadic computations}
\def\authorrunning{Tomas Petricek}

\newcommand{\sep}[0]{\; | \;}
\newcommand{\kvd}[1]{\textbf{#1}}  
\newcommand{\ident}[1]{\textit{#1}} 

\newcommand{\trans}[2]{ \llbracket #2 \rrbracket_{#1} }
\newcommand{\transs}[1]{ \llbracket #1 \rrbracket }
\newcommand{\transp}[2]{ \llbracket #2 \rrbracket_{#1} }

\begin{document}
\maketitle


\begin{abstract}
Monads have become a powerful tool for structuring effectful computations in functional
programming, because they make the order of effects explicit. When translating pure code
to a monadic version, we need to specify evaluation order explicitly. Two standard
translations give \emph{call-by-value} and \emph{call-by-name} semantics.
The resulting programs have different structure and types, which makes revisiting 
the choice difficult.

In this paper, we translate pure code to monadic using an additional operation 
\ident{malias} that abstracts out the evaluation strategy.  The \ident{malias} operation is 
based on \emph{computational comonads}; we use a categorical
framework to specify the laws that are required to hold about the operation. 

For any monad, we show implementations of \ident{malias} that give \emph{call-by-value} 
and \emph{call-by-name} semantics. Although we do not give \emph{call-by-need} semantics 
for \emph{all} monads, we show how to turn certain monads into an extended monad with
\emph{call-by-need} semantics, which partly answers an open question.
Moreover, using our unified translation, it is possible to change the evaluation strategy 
of functional code translated to the monadic form without changing its structure or types.
\end{abstract}


\section{Introduction}
Purely functional languages use lazy evaluation (also called \emph{call-by-need}) to allow
elegant programming with infinite data structures and to guarantee that a program will not 
evaluate a diverging term unless it is needed to obtain the final result. However, 
reasoning about lazy evaluation is difficult thus it is not suitable for programming with 
effects.

An elegant way to embed effectful computations in lazy functional languages, introduced
by Moggi \cite{monads-moggi} and Wadler \cite{monads-wadler}, is to use monads. Monads 
embed effects in a purely functional setting and explicitly specify the evaluation order
of monadic (effectful) operations.

Wadler \cite{monads-wadler} gives two ways of translating pure programs to a corresponding monadic
version. One approach leads to a \emph{call-by-value} semantics, where effects of function arguments
are performed before calling a function. However, if an argument has an effect and terminates the
program, this may not be appropriate if the function can successfully complete without 
using the argument. The second approach gives a \emph{call-by-name} semantics, where effects are 
performed only if the argument is actually used. However, this approach is also not always suitable, 
because an effect may be performed repeatedly. Wadler leaves an open question 
whether there is a translation that would correspond to \emph{call-by-need} semantics, 
where effects are performed only when the result is needed, but at most once.

The main contribution of this paper is an alternative translation of functional code to a 
monadic form, parameterized by an operation \ident{malias}. The translation has the following
properties:

\begin{itemize}
\item A single translation gives monadic code with either call-by-name or call-by-value 
  semantics, depending on the definition of \ident{malias} (Section~\ref{sec:abstracting}).
  When used in languages such as Haskell, it is possible to write code that is 
  parameterized by the evaluation strategy (Section~\ref{sec:practical-strategypolymorphic}). 

\item The translation can be used to construct monads that provide the \emph{call-by-need} 
  semantics (Section~\ref{sec:practical-cbn}), which partly answers the open question posed by Wadler.
  Furthermore, for some monads, it is possible to use \emph{parallel call-by-need} semantics, 
  where arguments are evaluated in parallel with the body of a function (Section~\ref{sec:practical-parallel-cbn}).

\item The \ident{malias} operation has solid foundations in category theory. It arises
  from augmenting a \emph{monad} structure with a \emph{computational semi-comonad} based on the
  same functor (Section~\ref{sec:theory}). We use this theory to define laws that should 
  be obeyed by \ident{malias} implementations (Section~\ref{sec:abstracting-malias}).
\end{itemize}

This paper was inspired by work on \emph{joinads} \cite{joinads-haskell11}, which 
introduced the \ident{malias} operation for a similar purpose. However, operations with
the same type and similar laws appear several times in the literature. We return to joinads 
in Section~\ref{sec:practical-joinads} and review other related work in 
Section~\ref{sec:related-work}.


\subsection{Translating to monadic code}
\label{sec:intro-translation}

We first demonstrate the two standard options for translating purely functional 
code to monadic form. Consider the following two functions that use
\ensuremath{\Varid{pureLookupInput}} to read some configuration property. Assuming the configuration is
already loaded in memory, we can write the following pure computation\footnote{Examples 
are written in Haskell and can be found at: \url{http://www.cl.cam.ac.uk/~tp322/papers/malias.html}}:

\begin{hscode}\SaveRestoreHook
\column{B}{@{}>{\hspre}l<{\hspost}@{}}%
\column{3}{@{}>{\hspre}l<{\hspost}@{}}%
\column{15}{@{}>{\hspre}l<{\hspost}@{}}%
\column{E}{@{}>{\hspre}l<{\hspost}@{}}%
\>[B]{}\Varid{chooseSize}\mathbin{::}\Conid{Int}\to \Conid{Int}\to \Conid{Int}{}\<[E]%
\\
\>[B]{}\Varid{chooseSize}\;\Varid{new}\;\Varid{legacy}\mathrel{=}{}\<[E]%
\\
\>[B]{}\hsindent{3}{}\<[3]%
\>[3]{}\mathbf{if}\;\Varid{new}\mathbin{>}\mathrm{0}\;\mathbf{then}\;\Varid{new}\;\mathbf{else}\;\Varid{legacy}{}\<[E]%
\\[\blanklineskip]%
\>[B]{}\Varid{resultSize}\mathbin{::}\Conid{Int}{}\<[E]%
\\
\>[B]{}\Varid{resultSize}\mathrel{=}{}\<[E]%
\\
\>[B]{}\hsindent{3}{}\<[3]%
\>[3]{}\Varid{chooseSize}\;{}\<[15]%
\>[15]{}(\Varid{pureLookupInput}\;\text{\tt \char34 new\char95 size\char34})\;{}\<[E]%
\\
\>[15]{}(\Varid{pureLookupInput}\;\text{\tt \char34 legacy\char95 size\char34}){}\<[E]%
\ColumnHook
\end{hscode}\resethooks
The \ensuremath{\Varid{resultSize}} function reads two different configuration keys and 
chooses one of them using \ensuremath{\Varid{chooseSize}}. When using a language with lazy evaluation, 
the call \ensuremath{\Varid{pureLookupInput}\;\text{\tt \char34 legacy\char95 size\char34}} is performed only when the value of 
\ensuremath{\text{\tt \char34 new\char95 size\char34}} is less than or equal to zero. 

To modify the function to actually read configuration from a file as opposed 
to performing in-memory lookup, we now use \ensuremath{\Varid{lookupInput}} which returns \ensuremath{\Conid{IO}\;\Conid{Int}}
instead of the \ensuremath{\Varid{pureLookupInput}} function. Then we need to modify the two above functions.
There are two mechanical ways that give different semantics.

\paragraph{Call-by-value.} In the first style, we call \ensuremath{\Varid{lookupInput}} and then 
apply monadic bind on the resulting computation. This reads both of the configuration
values before calling the \ensuremath{\Varid{chooseSize}} function, and so arguments are fully
evaluated before the body of a function as in the \emph{call-by-value} evaluation strategy:

\begin{hscode}\SaveRestoreHook
\column{B}{@{}>{\hspre}l<{\hspost}@{}}%
\column{3}{@{}>{\hspre}l<{\hspost}@{}}%
\column{E}{@{}>{\hspre}l<{\hspost}@{}}%
\>[B]{}\Varid{chooseSize}_{\textit{cbv}}\mathbin{::}\Conid{Int}\to \Conid{Int}\to \Conid{IO}\;\Conid{Int}{}\<[E]%
\\
\>[B]{}\Varid{chooseSize}_{\textit{cbv}}\;\Varid{new}\;\Varid{legacy}\mathrel{=}{}\<[E]%
\\
\>[B]{}\hsindent{3}{}\<[3]%
\>[3]{}\Varid{return}\;(\mathbf{if}\;\Varid{new}\mathbin{>}\mathrm{0}\;\mathbf{then}\;\Varid{new}\;\mathbf{else}\;\Varid{legacy}){}\<[E]%
\\[\blanklineskip]%
\>[B]{}\Varid{resultSize}_{\textit{cbv}}\mathbin{::}\Conid{IO}\;\Conid{Int}{}\<[E]%
\\
\>[B]{}\Varid{resultSize}_{\textit{cbv}}\mathrel{=}\mathbf{do}{}\<[E]%
\\
\>[B]{}\hsindent{3}{}\<[3]%
\>[3]{}\Varid{new}\leftarrow \Varid{lookupInput}_{\textit{cbv}}\;\text{\tt \char34 new\char95 size\char34}{}\<[E]%
\\
\>[B]{}\hsindent{3}{}\<[3]%
\>[3]{}\Varid{legacy}\leftarrow \Varid{lookupInput}_{\textit{cbv}}\;\text{\tt \char34 legacy\char95 size\char34}{}\<[E]%
\\
\>[B]{}\hsindent{3}{}\<[3]%
\>[3]{}\Varid{chooseSize}_{\textit{cbv}}\;\Varid{new}\;\Varid{legacy}{}\<[E]%
\ColumnHook
\end{hscode}\resethooks
In this version of the translation, a function of type \ensuremath{\Conid{A}\to \Conid{B}} is turned into 
a function \ensuremath{\Conid{A}\to \Conid{M}\;\Conid{B}}. For example, the \ensuremath{\Varid{chooseSize}_{\textit{cbv}}} function takes integers
as parameters and returns a computation that returns an integer and may perform 
some effects. When calling a function in this setting, the arguments may not be
fully evaluated (the functional part is still lazy), but the effects associated
with obtaining the value of the argument happen before the function call.

For example, if the call \ensuremath{\Varid{lookupInput}_{\textit{cbv}}\;\text{\tt \char34 new\char95 size\char34}} read a file and then returned 
1024, but the operation \ensuremath{\Varid{lookupInput}_{\textit{cbv}}\;\text{\tt \char34 legacy\char95 size\char34}} caused the program to crash
because a specified key was not present in a configuration file, then the entire 
program would crash.

\paragraph{Call-by-name.} In the second style, we pass unevaluated computations 
as arguments to functions. This means we call \ensuremath{\Varid{lookupInput}} to create an effectful 
computation that will read the input, but the computation is then passed to 
\ensuremath{\Varid{chooseSize}}, which may not need to evaluate it:

\begin{hscode}\SaveRestoreHook
\column{B}{@{}>{\hspre}l<{\hspost}@{}}%
\column{3}{@{}>{\hspre}l<{\hspost}@{}}%
\column{20}{@{}>{\hspre}l<{\hspost}@{}}%
\column{E}{@{}>{\hspre}l<{\hspost}@{}}%
\>[B]{}\Varid{chooseSize}_{\textit{cbn}}\mathbin{::}\Conid{IO}\;\Conid{Int}\to \Conid{IO}\;\Conid{Int}\to \Conid{IO}\;\Conid{Int}{}\<[E]%
\\
\>[B]{}\Varid{chooseSize}_{\textit{cbn}}\;\Varid{new}\;\Varid{legacy}\mathrel{=}\mathbf{do}{}\<[E]%
\\
\>[B]{}\hsindent{3}{}\<[3]%
\>[3]{}\Varid{newVal}\leftarrow \Varid{new}{}\<[E]%
\\
\>[B]{}\hsindent{3}{}\<[3]%
\>[3]{}\mathbf{if}\;\Varid{newVal}\mathbin{>}\mathrm{0}\;\mathbf{then}\;\Varid{new}\;\mathbf{else}\;\Varid{legacy}{}\<[E]%
\\[\blanklineskip]%
\>[B]{}\Varid{resultSize}_{\textit{cbn}}\mathbin{::}\Conid{IO}\;\Conid{Int}{}\<[E]%
\\
\>[B]{}\Varid{resultSize}_{\textit{cbn}}\mathrel{=}{}\<[E]%
\\
\>[B]{}\hsindent{3}{}\<[3]%
\>[3]{}\Varid{chooseSize}_{\textit{cbn}}\;{}\<[20]%
\>[20]{}(\Varid{lookupInput}_{\textit{cbn}}\;\text{\tt \char34 new\char95 size\char34})\;{}\<[E]%
\\
\>[20]{}(\Varid{lookupInput}_{\textit{cbn}}\;\text{\tt \char34 legacy\char95 size\char34}){}\<[E]%
\ColumnHook
\end{hscode}\resethooks
The translation turns a function of type \ensuremath{\Conid{A}\to \Conid{B}} into a function \ensuremath{\Conid{M}\;\Conid{A}\to \Conid{M}\;\Conid{B}}. 
This means that the \ensuremath{\Varid{chooseSize}_{\textit{cbn}}} function takes a computation that performs 
the I/O effect and reads information from the configuration file, as opposed to 
taking a value whose effects were already performed. 

Following the mechanical translation, \ensuremath{\Varid{chooseSize}_{\textit{cbn}}} returns a monadic computation that 
evaluates the first argument and then behaves either as \ensuremath{\Varid{new}} or as \ensuremath{\Varid{legacy}}, depending on the 
obtained value. When the resulting computation is executed, the computation which
reads the value of the \ensuremath{\text{\tt \char34 new\char95 size\char34}} key may be executed repeatedly. First, inside the 
\ensuremath{\Varid{chooseSize}_{\textit{cbn}}} function and then repeatedly when the result of this function is evaluated.
In this particular example, we can easily change the code to perform the effect just once, 
but this is not generally possible for computations obtained by the \emph{call-by-name} translation.


\section{Abstracting evaluation strategy}
\label{sec:abstracting}

The translations demonstrated in the previous section have two major problems. Firstly, 
it is not easy to switch between the two -- when we introduce effects using 
monads, we need to decide to use one or the other style and changing between them later
on involves rewriting of the program and changing types. Secondly, even in the \ensuremath{\Conid{IO}} monad,
we cannot easily implement a \emph{call-by-need} strategy that would perform effects only
when a value is needed, but at most once.

\subsection{Translation using aliasing}
\label{sec:abstracting-aliasing}

To solve these problems, we propose an alternative translation. We require a monad \ensuremath{\Varid{m}} with
an additional operation \ident{malias} that abstracts out the evaluation strategy and has
a type \ensuremath{\Varid{m}\;\Varid{a}\to \Varid{m}\;(\Varid{m}\;\Varid{a})}. The term \emph{aliasing} refers to the fact that some part
of effects may be performed once and their results shared in multiple monadic computations.
The translation of the previous example using \ident{malias} looks as follows:

\begin{hscode}\SaveRestoreHook
\column{B}{@{}>{\hspre}l<{\hspost}@{}}%
\column{3}{@{}>{\hspre}l<{\hspost}@{}}%
\column{E}{@{}>{\hspre}l<{\hspost}@{}}%
\>[B]{}\Varid{chooseSize}\mathbin{::}\Conid{IO}\;\Conid{Int}\to \Conid{IO}\;\Conid{Int}\to \Conid{IO}\;\Conid{Int}{}\<[E]%
\\
\>[B]{}\Varid{chooseSize}\;\Varid{new}\;\Varid{legacy}\mathrel{=}\mathbf{do}{}\<[E]%
\\
\>[B]{}\hsindent{3}{}\<[3]%
\>[3]{}\Varid{newVal}\leftarrow \Varid{new}{}\<[E]%
\\
\>[B]{}\hsindent{3}{}\<[3]%
\>[3]{}\mathbf{if}\;\Varid{newVal}\mathbin{>}\mathrm{0}\;\mathbf{then}\;\Varid{new}\;\mathbf{else}\;\Varid{legacy}{}\<[E]%
\\[\blanklineskip]%
\>[B]{}\Varid{resultSize}\mathbin{::}\Conid{IO}\;\Conid{Int}{}\<[E]%
\\
\>[B]{}\Varid{resultSize}\mathrel{=}\mathbf{do}{}\<[E]%
\\
\>[B]{}\hsindent{3}{}\<[3]%
\>[3]{}\Varid{new}\leftarrow \Varid{malias}\;(\Varid{lookupInput}\;\text{\tt \char34 new\char95 size\char34}){}\<[E]%
\\
\>[B]{}\hsindent{3}{}\<[3]%
\>[3]{}\Varid{legacy}\leftarrow \Varid{malias}\;(\Varid{lookupInput}\;\text{\tt \char34 legacy\char95 size\char34}){}\<[E]%
\\
\>[B]{}\hsindent{3}{}\<[3]%
\>[3]{}\Varid{chooseSize}\;\Varid{new}\;\Varid{legacy}{}\<[E]%
\ColumnHook
\end{hscode}\resethooks
The types of functions and access to function parameters are translated in the same 
way as in the \emph{call-by-name} translation. The \ensuremath{\Varid{chooseSize}} function returns a
computation \ensuremath{\Conid{IO}\;\Conid{Int}} and its parameters also become computations of type \ensuremath{\Conid{IO}\;\Conid{Int}}.
When using the value of a parameter, the computation is evaluated using monadic bind
(e.g.\ the line \ensuremath{\Varid{newVal}\leftarrow \Varid{new}} in \ensuremath{\Varid{chooseSize}}).

The computations passed as arguments are not the original computations as in the 
\emph{call-by-name} translation. The translation inserts a call to \ident{malias} for every 
function argument (and also for every let-bound value, which can be encoded using lambda abstraction 
and application). The computation returned by \ident{malias} has a type \ensuremath{\Varid{m}\;(\Varid{m}\;\Varid{a})},
which makes it possible to perform the effects at two different call sites:

\begin{itemize}
\item When simulating the \emph{call-by-value} strategy, all effects are performed when binding
the outer monadic computation before a function call. 

\item When simulating the \emph{call-by-name} strategy, all effects are performed when binding
the inner monadic computation, when the value is actually needed.
\end{itemize}

These two strategies can be implemented by two simple definitions of \ident{malias}. However,
by delegating the implementation of \ident{malias} to the monad, we make it possible to 
implement more advanced strategies as well. We discuss some of them later in Section~\ref{sec:practical}.
We keep the translation informal until Section~\ref{sec:abstracting-translation} and 
discuss the \ident{malias} operation in more detail first.


\paragraph{Implementing call-by-name.}
To implement the \emph{call-by-name} strategy, the \ident{malias} operation needs to 
return the computation specified as an argument inside the monad. In the type \ensuremath{\Varid{m}\;(\Varid{m}\;\Varid{a})},
the outer \ensuremath{\Varid{m}} will not carry any effects and the inner \ensuremath{\Varid{m}} will be the same as the original
computation:

\begin{hscode}\SaveRestoreHook
\column{B}{@{}>{\hspre}l<{\hspost}@{}}%
\column{E}{@{}>{\hspre}l<{\hspost}@{}}%
\>[B]{}\Varid{malias}\mathbin{::}\Varid{m}\;\Varid{a}\to \Varid{m}\;(\Varid{m}\;\Varid{a}){}\<[E]%
\\
\>[B]{}\Varid{malias}\;\Varid{m}\mathrel{=}\Varid{return}\;\Varid{m}{}\<[E]%
\ColumnHook
\end{hscode}\resethooks
From the monad laws (see Figure~\ref{fig:monads-alternatives}), we know that applying monadic
bind to a computation created from a value using \ensuremath{\Varid{return}} is equivalent to just passing the value 
to the rest of the computation. This means that the additional binding in the translation does not 
have any effect and the resulting program behaves as the \emph{call-by-name} strategy. A 
complete proof can be found in Appendix~\ref{sec:appendix-cbv-cbn-equivalence}.

\paragraph{Implementing call-by-value.}
Implementing the \emph{call-by-value} strategy is similarly simple. In the returned computation
of type \ensuremath{\Varid{m}\;(\Varid{m}\;\Varid{a})}, the computation corresponding to the outer \ensuremath{\Varid{m}} needs to perform all the effects.
The computation corresponding to the inner \ensuremath{\Varid{m}} will be a computation that simply returns the 
previously computed value without performing any effects:

\begin{hscode}\SaveRestoreHook
\column{B}{@{}>{\hspre}l<{\hspost}@{}}%
\column{E}{@{}>{\hspre}l<{\hspost}@{}}%
\>[B]{}\Varid{malias}\mathbin{::}\Varid{m}\;\Varid{a}\to \Varid{m}\;(\Varid{m}\;\Varid{a}){}\<[E]%
\\
\>[B]{}\Varid{malias}\;\Varid{m}\mathrel{=}\Varid{m}\bind (\Varid{return}\mathbin{\circ}\Varid{return}){}\<[E]%
\ColumnHook
\end{hscode}\resethooks
In Haskell, the second line could be written as \ensuremath{\Varid{liftM}\;\Varid{return}\;\Varid{m}}. The \ensuremath{\Varid{liftM}} operation
represents the functor associated with the monad. This means that binding on the returned 
computation performs all the effects, obtains a value \ensuremath{\Varid{v}} and returns a computation 
\ensuremath{\Varid{return}\;\Varid{v}}. 

When calling a function that takes an argument of type \ensuremath{\Varid{m}\;\Varid{a}}, the argument passed to it 
using this implementation of \ident{malias} will always be constructed using the \ensuremath{\Varid{return}}
operation. Hence the resulting behaviour is equivalent to the original \emph{call-by-value} 
translation. Detailed proof can be found in Appendix~\ref{sec:appendix-cbv-cbn-equivalence}.


\begin{figure}

\begin{multicols}{2}
Functor with \emph{unit} and \emph{join}:
\vspace{-0.8em}
\begin{hscode}\SaveRestoreHook
\column{B}{@{}>{\hspre}l<{\hspost}@{}}%
\column{9}{@{}>{\hspre}l<{\hspost}@{}}%
\column{E}{@{}>{\hspre}l<{\hspost}@{}}%
\>[B]{}\Varid{unit}{}\<[9]%
\>[9]{}\mathbin{::}\Varid{a}\to \Varid{m}\;\Varid{a}{}\<[E]%
\\
\>[B]{}\Varid{map}{}\<[9]%
\>[9]{}\mathbin{::}\Varid{m}\;\Varid{a}\to (\Varid{a}\to \Varid{b})\to \Varid{m}\;\Varid{b}{}\<[E]%
\\
\>[B]{}\Varid{join}{}\<[9]%
\>[9]{}\mathbin{::}\Varid{m}\;(\Varid{m}\;\Varid{a})\to \Varid{m}\;\Varid{a}{}\<[E]%
\ColumnHook
\end{hscode}\resethooks
\emph{Join} laws (\emph{unit} and \emph{map} laws omitted):
\vspace{-0.7em}
\begin{hscode}\SaveRestoreHook
\column{B}{@{}>{\hspre}l<{\hspost}@{}}%
\column{E}{@{}>{\hspre}l<{\hspost}@{}}%
\>[B]{}\Varid{join}\mathbin{\circ}\Varid{map}\;\Varid{join}\mathrel{=}\Varid{join}\mathbin{\circ}\Varid{join}{}\<[E]%
\\
\>[B]{}\Varid{join}\mathbin{\circ}\Varid{map}\;\Varid{unit}\mathrel{=}\Varid{id}\mathrel{=}\Varid{join}\mathbin{\circ}\Varid{unit}{}\<[E]%
\\
\>[B]{}\Varid{join}\mathbin{\circ}\Varid{map}\;(\Varid{map}\;\Varid{f})\mathrel{=}\Varid{map}\;\Varid{f}\mathbin{\circ}\Varid{join}{}\<[E]%
\ColumnHook
\end{hscode}\resethooks

Definition using \emph{bind} (\ensuremath{\bind }) and \emph{return}:
\vspace{-0.8em}
\begin{hscode}\SaveRestoreHook
\column{B}{@{}>{\hspre}c<{\hspost}@{}}%
\column{BE}{@{}l@{}}%
\column{9}{@{}>{\hspre}l<{\hspost}@{}}%
\column{E}{@{}>{\hspre}l<{\hspost}@{}}%
\>[B]{}\Varid{return}{}\<[BE]%
\>[9]{}\mathbin{::}\Varid{a}\to \Varid{m}\;\Varid{a}{}\<[E]%
\\
\>[B]{}\bind {}\<[BE]%
\>[9]{}\mathbin{::}\Varid{m}\;\Varid{a}\to (\Varid{a}\to \Varid{m}\;\Varid{b})\to \Varid{m}\;\Varid{b}{}\<[E]%
\ColumnHook
\end{hscode}\resethooks
\vspace{0em}

Monad laws about \emph{bind} and \emph{return}:
\vspace{-0.8em}
\begin{hscode}\SaveRestoreHook
\column{B}{@{}>{\hspre}l<{\hspost}@{}}%
\column{E}{@{}>{\hspre}l<{\hspost}@{}}%
\>[B]{}\Varid{return}\;\Varid{a}\bind \Varid{f}\equiv \Varid{f}\;\Varid{a}{}\<[E]%
\\
\>[B]{}\Varid{m}\bind \Varid{return}\equiv \Varid{m}{}\<[E]%
\\
\>[B]{}(\Varid{m}\bind \Varid{f})\bind \Varid{g}\equiv \Varid{m}\bind (\lambda \Varid{x}\to \Varid{f}\;\Varid{x}\bind \Varid{g}){}\<[E]%
\ColumnHook
\end{hscode}\resethooks
\end{multicols}
\vspace{-0.8em}
\caption{Two equivalent ways of defining monads with monad laws}
\label{fig:monads-alternatives}
\end{figure}


\subsection{The malias operation laws}
\label{sec:abstracting-malias}

In order to define a reasonable evaluation strategy, we require the \ident{malias} 
operation to obey a number of laws. The laws follow from the theoretical background that
is discussed in Section~\ref{sec:theory}, namely from the fact that \ident{malias} is the
\emph{cojoin} operation of a computational semi-comonad.

The laws that relate \ident{malias} to the monad are easier to write in terms of \ensuremath{\Varid{join}},
\ensuremath{\Varid{map}} and \ensuremath{\Varid{unit}} than using the formulation based on \ensuremath{\bind } and \ensuremath{\Varid{return}}. For completeness,
the two equivalent definitions of monads with the monad laws are shown in Figure~\ref{fig:monads-alternatives}. 
Although we do not show it, one can be easily defined in terms of the other.
The required laws for \ensuremath{\Varid{malias}} are the following:

\[
\begin{array}{rclcl}
\textit{map}~(\textit{map}~f) \circ \textit{malias} &=& \textit{malias} \circ (\textit{map}~f) 
  &\quad\quad&(\textit{naturality}) \\
 \textit{map}~\textit{malias} \circ \textit{malias} &=& \textit{malias} \circ \textit{malias}
  &&(\textit{associativity}) \\ 
              \textit{malias} \circ \textit{unit} &=& \textit{unit} \circ \textit{unit}
  &&(\textit{computationality}) \\              
                \textit{join} \circ \textit{malias} &=& \textit{id}
  &&(\textit{identity}) \\
\end{array}
\]
The first two laws follow from the fact that \ident{malias} is a \emph{cojoin} operation 
of a comonad. The \emph{naturality} law specifies that applying a function to a value inside 
a computation is the same as applying the function to a value inside an aliased computation.
The \emph{associativity} law specifies that aliasing an aliased computation is the same
as aliasing a computation produced by an aliased computation.

The \emph{computationality} law is derived from the fact that the comonad defining 
\ident{malias} is a \emph{computational comonad} with \emph{unit} as one of the components. 
The law specifies that aliasing of a pure computation creates a pure computation.
Finally, the \emph{identity} law relates \ensuremath{\Varid{malias}} with the monadic structure, by requiring 
that \ensuremath{\Varid{join}} is a \emph{left inverse} of \ensuremath{\Varid{malias}}. Intuitively, it specifies that aliasing a 
computation of type \ensuremath{\Varid{m}\;\Varid{a}} and then joining the result returns the original computation. 

All four laws hold for the two implementations of \ident{malias} presented in 
the previous section. We prove that the laws hold for any monad using the standard monad
laws. The proofs can be found in Appendix~\ref{sec:appendix-cbv-cbn-proofs}.
We discuss the intuition behind the laws in Section~\ref{sec:abstracting-intuition} and 
describe their categorical foundations in Section~\ref{sec:theory}. The next section 
formally presents the translation algorithm.


\begin{figure}
\[
\begin{array}{lcl}
\transp{cbn}{x}                &=& x\\
\transp{cbn}{\lambda x . e}    &=& \ident{unit}~(\lambda x . \trans{cbn}{e})\\
\transp{cbn}{e_1 \: e_2}       &=& \ident{bind}~\trans{cbn}{e_1}~(\lambda f . f~\trans{cbn}{e_2})\quad\quad\quad\quad\quad\quad  \\ 
\transp{cbn}{\kvd{let} \: x = e_1 \: \kvd{in} \: e_2}   &=&  (\lambda x . \trans{cbn}{e_2})~\trans{cbn}{e_1}  \\
\end{array}
\]
\vspace{-1em}
\caption{Wadler's call-by-name translation of $\lambda$ calculus.}
\label{fig:translation-cbn}
\end{figure}

\begin{figure}
\[
\begin{array}{lcl}
\transp{cbv}{x}                &=& \ident{unit}~x\\
\transp{cbv}{\lambda x . e}    &=& \ident{unit}~(\lambda x . \trans{cbv}{e})\\
\transp{cbv}{e_1 \: e_2}       &=& \ident{bind}~\trans{cbv}{e_1}~(\lambda f . ~\ident{bind}~\trans{cbv}{e_2}~(\lambda x . f~x))  \\ 
\transp{cbv}{\kvd{let} \: x = e_1 \: \kvd{in} \: e_2}   &=&  \ident{bind}~\trans{cbv}{e_1}~(\lambda x . \trans{cbv}{e_2})
\end{array}
\]
\vspace{-1em}
\caption{Wadler's call-by-value translation of $\lambda$ calculus.}
\label{fig:translation-cbv}
\end{figure}


\subsection{Lambda calculus translation}
\label{sec:abstracting-translation}

The \emph{call-by-name} and \emph{call-by-value} translations given in 
Section~\ref{sec:intro-translation} were first formally introduced by Wadler \cite{monads-wadler}.
In this section, we present a similar formal definition of our translation based on the
\ident{malias} operation. For our source language, we use a simply-typed $\lambda$ calculus with 
let-binding:

\begin{equation*}
\begin{array}{lclccl}
e &\in& \textit{Expr} & e & ::= & x \sep \lambda x . e \sep e_1 \: e_2 \sep \kvd{let} \: x = e_1 \: \kvd{in} \: e_2 \\
\tau   &\in& \textit{Type}    &\tau &::=& \alpha \sep \tau_1 \rightarrow \tau_2
\end{array}
\end{equation*}
The target language of the translation is identical but for one exception -- it adds a type scheme $M~\tau$
representing monadic computations. The \emph{call-by-name} and \emph{call-by-value} translation of the lambda calculus 
are shown in Figure~\ref{fig:translation-cbn} and Figure~\ref{fig:translation-cbv}, respectively. In the
translation, we write \ensuremath{\bind } as \ensuremath{\Varid{bind}}. The translation of types and typing 
judgements are omitted for simplicity and can be found in the original paper \cite{monads-wadler}.

Our translation, called \emph{call-by-alias}, is presented below. The translation has 
similar structure to Wadler's \emph{call-by-name} translation, but it inserts \ident{malias} operation
in the last two cases:

\[
\begin{array}{lcl}
\transp{cba}{\alpha}           &=& \alpha\\
\transp{cba}{\tau_1 \rightarrow \tau_2} &=& M~\transp{cba}{\tau_1} \rightarrow M~\transp{cba}{\tau_2}\\
\\[-0.4em]
\transp{cba}{x}                &=& x\\
\transp{cba}{\lambda x . e}    &=& \ident{unit}~(\lambda x . \trans{cba}{e})\\
\transp{cba}{e_1 \: e_2}       &=& \ident{bind}~\trans{cba}{e_1}~(\lambda f . \ident{bind}~(\ident{malias}~\trans{cba}{e_2})~f)  \\ 
\transp{cba}{\kvd{let} \: x = e_1 \: \kvd{in} \: e_2}   &=&  \ident{bind}~(\ident{malias}~\trans{cba}{e_1})~(\lambda x . \trans{cba}{e_2})  \\
\end{array}
\]
\vspace{0.0em}
\[
\transp{cba}{x_1:\tau_1, \ldots, x_n:\tau_n \vdash e : \tau} =
x_1:M~\transp{cba}{\tau_1}, \ldots, x_n:M~\transp{cba}{\tau_n} \vdash \transp{cba}{e} : M~\transp{cba}{\tau}
\]
The translation turns user-defined variables of type $\tau$ into variables of type
$M~\tau$. A variable access $x$ is translated to a variable access, which 
now represents a computation (that may have effects). A lambda expression
is turned into a lambda expression wrapped in a pure monadic computation.

The two interesting cases are application and let-binding. When translating function
application, we bind on the computation representing the function. We want to call the 
function $f$ with an aliased computation as an argument. This is achieved by passing 
the translated argument to \ident{malias} and then applying \ident{bind} again.
The translation of let-binding is similar, but slightly simpler, because it does not
need to use \ident{bind} to obtain a function. 

The definition of $\transp{cba}{-}$ includes a well-typedness-preserving translation of
typing judgements. The details of the proof can be found in Appendix~\ref{sec:appendix-preserve-typing}.
In the translation, let-binding is equivalent to $(\lambda x.e_2)~e_1$, but we include 
it to aid the intuition and to simplify motivating examples in the next section.


\subsection{The meaning of malias laws}
\label{sec:abstracting-intuition}
Having provided the translation, we can discuss the intuition behind the \ensuremath{\Varid{malias}} laws
and what they imply about the translated code. The \emph{naturality} law specifies that 
\ensuremath{\Varid{malias}} is a natural transformation. Although we do not give a formal proof, we argue that
the law follows from the parametricity of the \ensuremath{\Varid{malias}} type signature and can be obtained 
using the method described by Voigtl\"ander \cite{types-free-theorems-classes}.

\paragraph{Effect conservation.}
The meaning of \emph{associativity} and \emph{identity} can be informally demonstrated by 
treating effects as units of information. Given a computation type involving a number of occurrences of
\ensuremath{\Varid{m}}, we say that there are \emph{effects} (or \emph{information}) associated with each occurrence 
of \ensuremath{\Varid{m}}.

The \emph{identity} law specifies that \ensuremath{\Varid{join}\mathbin{\circ}\Varid{malias}} does not lose effects -- given a computation 
of type \ensuremath{\Varid{m}\;\Varid{a}}, the \ensuremath{\Varid{malias}} operation constructs a computation of type \ensuremath{\Varid{m}\;(\Varid{m}\;\Varid{a})}. This is done by 
splitting the effects of the computation between two monadic computations. Requiring that applying 
\ensuremath{\Varid{join}} to the new computation returns the original computation means that all the effects of 
the original computation are preserved, because \ensuremath{\Varid{join}} combines the effects of the two computations.

The \ensuremath{\Varid{associativity}} law specifies how the effects should be split. When applying \ensuremath{\Varid{malias}} to
an aliased computation of type \ensuremath{\Varid{m}\;(\Varid{m}\;\Varid{a})}, we can apply the operation to the inner or the outer
\ensuremath{\Varid{m}}. Underlining second aliasing, the following two computations should be 
equal: $\underline{m~(m}~(m~a)) = m~(\underline{m~(m}~a))$. The law forbids implementations that
split the effects in an asymmetric way. For instance, if \ensuremath{\Varid{m}\;\Varid{a}} represents a step-wise
computation that can be executed by a certain number of steps, \ensuremath{\Varid{malias}} cannot execute the first half 
of steps and return a computation that evaluates the other half. The proportions associated
with individual \ensuremath{\Varid{m}} values would be $\nicefrac{1}{4}$, $\nicefrac{1}{4}$ and $\nicefrac{1}{2}$
on the left and $\nicefrac{1}{2}$, $\nicefrac{1}{4}$ and $\nicefrac{1}{4}$ on the right-hand side.

\paragraph{Semantics-preserving transformations.}
The \emph{computationality} and \emph{identity} (again) laws also specify that certain 
semantics-preserving transformations on the original source code correspond to equivalent
terms in the code translated using the \emph{call-by-alias} transformation. The source 
transformations corresponding to \emph{computationality} and \emph{identity}, respectively, 
are the following:

\[
\begin{array}{rcl}
\textbf{let}~f = \lambda x.e_1~\textbf{in}~e_2 & \equiv & e_2 [f \leftarrow \lambda x.e_1]
\\[0.3em]
\textbf{let}~x = e~\textbf{in}~x & \equiv & e
\end{array}
\]
The construction of the lambda function in the first equation is the only place where the translation 
inserts \ensuremath{\Varid{unit}}. The \emph{computationality} law can be applied when a lambda abstraction appears in a 
position where \ensuremath{\Varid{malias}} is inserted. Thanks to the first monad law (Figure~\ref{fig:monads-alternatives}),
the value assigned to \ensuremath{\Varid{f}} in the translation is $\textit{unit}~(\lambda x.e_1)$, which is 
equivalent to the translation of a term after the substitution. In the second equation,
the left-hand side is translated as \ensuremath{\Varid{bind}\;(\Varid{malias}\;\Varid{e})\;\Varid{id}}, which is equivalent to 
\ensuremath{\Varid{join}\;(\Varid{malias}\;\Varid{e})}, so the equation directly corresponds to the \emph{identity} law.

\paragraph{Discussion of completeness.}
The above discussion, together with the theoretical foundations introduced in the next 
section, supports the claim that our laws are necessary. We do not argue that our set of laws 
is complete -- for instance, we might want to specify that aliasing of an already aliased 
computation has no effect, which is difficult to express in an equational form.

However, the definition of completeness, in this context, is elusive. One possible 
approach that we plan to investigate in future work is to expand the set of semantics-preserving
transformations that should hold, regardless of an evaluation strategy.


\newcommand{\catc}{\mathcal{C}}
\newcommand{\idf}[1]{ {\textnormal{\sffamily id}}_{#1} }

\section{Computational semi-bimonads}
\label{sec:theory}

In this section, we formally describe the structure that underlies a monad having a 
\ident{malias} operation as described in the previous section. Since the \ident{malias} 
operation corresponds to an operation of a comonad associated with a monad, we first review the 
definitions of a monad and a comonad. Monads are well-known structures in functional programming. 
Comonads are dual structures to monads that are less widespread, but they have also been used in 
functional programming and semantics (Section~\ref{sec:related-work}):

\vspace{0.5em}
\begin{definition}
A \emph{monad} over a category $\catc$ is a triple $(T, \eta, \mu)$
where $T : \catc \rightarrow \catc$ is a functor, $\eta : I_\catc \rightarrow T$ is
a natural transformation from the identity functor to $T$, and $\mu : T^2 \rightarrow T$
is a natural transformation, such that the following associativity and
identity conditions hold, for every object $A$:

\vspace{0.5em}
\begin{compactitem}
\item $\mu_{A} \circ T \mu_A = \mu_{A} \circ \mu_{T A}$
\item $\mu_{A} \circ \eta_{T A} = \idf{T A} = \mu_{A} \circ T \eta_{A}$
\end{compactitem}
\vspace{0.8em}
\end{definition}

\begin{definition}
\label{def:comonad}
A \emph{comonad} over a category $\catc$ is a triple $(T, \epsilon, \delta)$
where $T : \catc \rightarrow \catc$ is a functor, $\epsilon : T \rightarrow I_\catc$ is
a natural transformation from $T$ to the identity functor, and $\delta : T \rightarrow T^2$
is a natural transformation from $T$ to $T^2$, such that the following associativity and
identity conditions hold, for every object $A$:

\vspace{0.5em}
\begin{compactitem}
\item $T \delta_A \circ \delta_A = \delta_{T A} \circ \delta_A$
\item $\epsilon_{T A} \circ \delta_A = \idf{T A} = T \epsilon_A \circ \delta_A$
\end{compactitem}
\vspace{0.8em}
\end{definition}

In functional programming terms, the natural transformation $\eta$ corresponds to 
\ensuremath{\Varid{unit}\mathbin{::}\Varid{a}\to \Varid{m}\;\Varid{a}} and the natural transformation $\mu$ corresponds to \ensuremath{\Varid{join}\mathbin{::}\Varid{m}\;(\Varid{m}\;\Varid{a})\to \Varid{m}\;\Varid{a}}.
A comonad is a dual structure to a monad -- the natural transformation $\epsilon$ 
corresponds to an operation \ensuremath{\Varid{counit}\mathbin{::}\Varid{m}\;\Varid{a}\to \Varid{a}} and $\delta$ corresponds to 
\ensuremath{\Varid{cojoin}\mathbin{::}\Varid{m}\;\Varid{a}\to \Varid{m}\;(\Varid{m}\;\Varid{a})}. An equivalent formulation of comonads in functional programming 
uses an operation \ensuremath{\Varid{cobind}\mathbin{::}\Varid{m}\;\Varid{a}\to (\Varid{m}\;\Varid{a}\to \Varid{b})\to \Varid{m}\;\Varid{b}}, which is dual to \ensuremath{\bind } of monads.

A simple example of a comonad is the product comonad. The type \ensuremath{\Varid{m}\;\Varid{a}} stores the value of
\ensuremath{\Varid{a}} and some additional state \ensuremath{\Conid{S}}, meaning that $T A = A \times S$. The $\epsilon$ (or
\ensuremath{\Varid{counit}}) operation extracts the value $A$ ignoring the additional state. The $\delta$ 
(or \ensuremath{\Varid{cojoin}}) operation duplicates the state. In functional programming, the product 
comonad is equivalent to the reader monad $T A = S \rightarrow A$.

In this paper, we use a special variant of comonads. Computational comonads, introduced by
Brookes and Geva \cite{comonads-computational}, have an additional operation $\gamma$
together with laws specifying its properties:

\vspace{0.5em}
\begin{definition}
A computational comonad over a category $\catc$ is a quadruple $(T, \epsilon, \delta, \gamma)$
where $(T, \epsilon, \delta)$ is a comonad over $\catc$ and $\gamma : I_\catc \rightarrow T$
is a natural transformation such that, for every object $A$,

\vspace{0.5em}
\begin{compactitem}
\item $\epsilon_A \circ \gamma_A = \idf{A}$
\item $\delta_A \circ \gamma_A = \gamma_{T A} \circ \gamma_A.$
\end{compactitem}
\vspace{0.8em}
\end{definition}

A \emph{computational comonad} has an additional operation $\gamma$ which has the same type
as the $\eta$ operation of a monad, that is \ensuremath{\Varid{a}\to \Varid{m}\;\Varid{a}}. In the work on computational comonads,
the transformation $\gamma$ turns an extensional specification into an intensional specification
without additional computational information.

In our work, we do not need the natural transformation corresponding to \ensuremath{\Varid{counit}\mathbin{::}\Varid{m}\;\Varid{a}\to \Varid{a}}. 
We define a computational \emph{semi}-comonad, which is a computational comonad without the 
natural transformation $\epsilon$ and without the associated laws. The remaining structure is 
preserved:

\vspace{0.5em}
\begin{definition}
A \emph{computational semi-comonad} over a category $\catc$ is a triple $(T, \delta, \gamma)$
where $T : \catc \rightarrow \catc$ is a functor, $\delta : T \rightarrow T^2$
is a natural transformation from $T$ to $T^2$ and $\gamma : I_\catc \rightarrow T$ is 
a natural transformation from the identity functor to $T$, such that the following 
associativity and computationality conditions hold, for every object $A$:

\vspace{0.5em}
\begin{compactitem}
\item $T \delta_A \circ \delta_A = \delta_{T A} \circ \delta_A$
\item $\delta_A \circ \gamma_A = \gamma_{T A} \circ \gamma_A.$
\end{compactitem}
\vspace{0.8em}
\end{definition}

Finally, to define a structure that models our monadic computations with the \ident{malias}
operation, we combine the definition of a monad and computational semi-comonad.
We require that the two structures share the functor $T$ and that the natural transformation
$\eta : I_\catc \rightarrow T$ of a monad coincides with the natural transformation
$\gamma : I_\catc \rightarrow T$ of a computational comonad. 

\vspace{0.5em}
\begin{definition}
\label{def:comp-semi-bimon}
A \emph{computational semi-bimonad} over a category $\catc$ is a quadruple $(T, \eta, \mu, \delta)$
where $(T, \eta, \mu)$ is a monad over a category $\catc$ and $(T, \delta, \eta)$
is a computational semi-comonad over $\catc$, such that the following additional condition
holds, for every object $A$:

\vspace{0.5em}
\begin{compactitem}
\item $\mu_A \circ \delta_A = \idf{T A}$
\end{compactitem}
\vspace{0.8em}
\end{definition}

The definition of computational semi-bimonad relates the monadic and comonadic parts of 
the structure using an additional law. Given an object $A$, the law specifies that
taking $T A$ to $T^2 A$ using the natural transformation $\delta_A$ of a comonad and then 
back to $T A$ using the natural transformation $\mu_A$ is identity.

\subsection{Revisiting the laws}
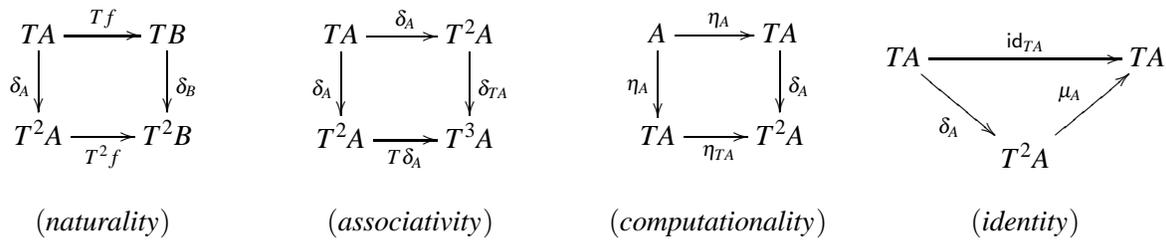
\begin{figure}
\begin{multicols}{4}
\[
\hspace{-2em}
\xymatrix{
    T A\ar[r]^{T f}\ar[d]_{\delta_A}    &    T B\ar[d]^{\delta_B}\\
    T^2 A\ar[r]_{T^2 f}                 &    T^2 B
}
\]
\[\hspace{-2em}
 (\textit{naturality}) \]

\[
\hspace{-2em}
\xymatrix{
    T A\ar[r]^{\delta_A}\ar[d]_{\delta_A}    &    T^2 A\ar[d]^{\delta_{T A}}\\
    T^2 A\ar[r]_{T \delta_{A}}               &    T^3 A
}
\]
\[\hspace{-2em}
 (\textit{associativity}) \]

\[
\hspace{-2em}
\xymatrix{
    A\ar[r]^{\eta_A}\ar[d]_{\eta_A}    &    T A\ar[d]^{\delta_{A}}\\
    T A\ar[r]_{\eta_{T A}}             &    T^2 A
}
\]
\[\hspace{-2em}
 (\textit{computationality}) \]

\[
\hspace{-2em}
\xymatrix{
    T A\ar[dr]_{ \delta_A }\ar[rr]^{ \idf{T A} }   &&   T A\\
    &T^2 A\ar[ur]^{\mu_A}
}
\]
\[\hspace{-2em}
 (\textit{identity}) \]
\end{multicols}

\caption{Diagramatic representation of the four additional properties of semi-bimonads}
\label{fig:computational-semi-bimonad}
\end{figure}

The laws of computational semi-bimonad as defined in the previous section are exactly the laws
of our monad equipped with the \ident{malias} operation. In this section, we briefly review the
laws and present the category theoretic version of all the laws demonstrated in 
Section~\ref{sec:abstracting-malias}. We require four laws in addition to the standard monad laws 
(which are omitted in the summary below). A diagrammatic demonstration is shown in 
Figure~\ref{fig:computational-semi-bimonad}. For all objects $A$ and $B$ of $\catc$ and for 
all $f : A \rightarrow B$ in $\catc$:

\[
\begin{array}{rclrl}
T^2 f \circ \delta_{A} &=& \delta_B \circ T f
&&            (\textit{naturality})\\
T \delta_A \circ \delta_A &=& \delta_{T A} \circ \delta_A
&\quad\quad&  (\textit{associativity})\\
\delta_A \circ \eta_A &=& \eta_{T A} \circ \eta_A
&&            (\textit{computationality})\\
\mu_A \circ \delta_A  &=& \idf{T A}
&&            (\textit{identity})\\
\end{array}
\]

The \emph{naturality} law follows from the fact that $\delta$ is a natural transformation and so
we did not state it explicitly in Definition~\ref{def:comp-semi-bimon}. However, it is one of the laws that are translated to
the functional programming interpretation. The \emph{associativity} law is a law of comonad -- the
other law in Definition~\ref{def:comonad} does not apply in our scenario, because we only work with \emph{semi}-comonad that 
does not have natural transformation $\epsilon$ (\ensuremath{\Varid{counit}}). The \emph{computationality} law 
is a law of a computational comonad and finally, the \emph{identity} law is the additional law
of \emph{computational semi-bimonads}.

\section{Abstracting evaluation strategy in practice}
\label{sec:practical}

In this section, we present several practical uses of the \ident{malias} operation.
We start by showing how to write monadic code that is parameterized over the evaluation
strategy and then consider expressing \emph{call-by-need} in this framework. Then we
also briefly consider \emph{parallel call-by-need} and the relation between \ident{malias}
and \emph{joinads} and the \ensuremath{\mathbf{docase}} notation \cite{joinads-haskell11}.


\subsection{Parameterization by evaluation strategy}
\label{sec:practical-strategypolymorphic}

One of the motivations of this work is that the standard monadic translations for 
\emph{call-by-name} and \emph{call-by-value} produce code with different structure. 
Section~\ref{sec:abstracting-translation} gave a translation that can be used with both of the evaluation 
strategies just by changing the definition of the \ident{malias} operation. In this 
section, we make one more step -- we show how to write code parameterized by evaluation
strategy.

We define a \emph{monad transformer} \cite{monad-transformer-interpreters} that takes
a monad and turns it into a monad with \ensuremath{\Varid{malias}} that implements a specific evaluation 
strategy. Our example can then be implemented using functions that are polymorphic over
the monad transformer. We continue using the previous example based on the \ensuremath{\Conid{IO}} monad, 
but the transformer can operate on any monad.

As a first step, we define a type class named \ensuremath{\Conid{MonadAlias}} that extends \ensuremath{\Conid{Monad}} with 
the \ident{malias} operation. To keep the code simple, we do not include comments 
documenting the laws:

\begin{hscode}\SaveRestoreHook
\column{B}{@{}>{\hspre}l<{\hspost}@{}}%
\column{3}{@{}>{\hspre}l<{\hspost}@{}}%
\column{E}{@{}>{\hspre}l<{\hspost}@{}}%
\>[B]{}\mathbf{class}\;\Conid{Monad}\;\Varid{m}\Rightarrow \Conid{MonadAlias}\;\Varid{m}\;\mathbf{where}{}\<[E]%
\\
\>[B]{}\hsindent{3}{}\<[3]%
\>[3]{}\Varid{malias}\mathbin{::}\Varid{m}\;\Varid{a}\to \Varid{m}\;(\Varid{m}\;\Varid{a}){}\<[E]%
\ColumnHook
\end{hscode}\resethooks
Next, we define two new types that represent monadic computations using the 
\emph{call-by-name} and \emph{call-by-value} evaluation strategy. The two types are
wrappers that make it possible to implement two different instances of \ensuremath{\Conid{MonadAlias}}
for any underlying monadic computation \ensuremath{\Varid{m}\;\Varid{a}}:

\begin{hscode}\SaveRestoreHook
\column{B}{@{}>{\hspre}l<{\hspost}@{}}%
\column{E}{@{}>{\hspre}l<{\hspost}@{}}%
\>[B]{}\mathbf{newtype}\;\Conid{CbV}\;\Varid{m}\;\Varid{a}\mathrel{=}\Conid{CbV}\;\{\mskip1.5mu \Varid{runCbV}\mathbin{::}\Varid{m}\;\Varid{a}\mskip1.5mu\}{}\<[E]%
\\
\>[B]{}\mathbf{newtype}\;\Conid{CbN}\;\Varid{m}\;\Varid{a}\mathrel{=}\Conid{CbN}\;\{\mskip1.5mu \Varid{runCbN}\mathbin{::}\Varid{m}\;\Varid{a}\mskip1.5mu\}{}\<[E]%
\ColumnHook
\end{hscode}\resethooks
The snippet defines types \ensuremath{\Conid{CbV}} and \ensuremath{\Conid{CbN}} that represent two evaluation strategies. 
Figure~\ref{fig:monadalias-transformers} shows the implementation of three type classes
for these two types. The implementation of the \ensuremath{\Conid{Monad}} type class is the same for both
types, because it simply uses \ensuremath{\Varid{return}} and \ensuremath{\bind } operations of the underlying
monad. The implementation of \ensuremath{\Conid{MonadTrans}} wraps a monadic computation \ensuremath{\Varid{m}\;\Varid{a}} into a 
type \ensuremath{\Conid{CbV}\;\Varid{m}\;\Varid{a}} and \ensuremath{\Conid{CbN}\;\Varid{m}\;\Varid{a}}, respectively. Finally, the instances of the
\ensuremath{\Conid{MonadAlias}} type class associate the two implementations of \ident{malias} (from
Section~\ref{sec:abstracting-aliasing}) with the two data types.

\begin{figure} 
\begin{hscode}\SaveRestoreHook
\column{B}{@{}>{\hspre}l<{\hspost}@{}}%
\column{3}{@{}>{\hspre}l<{\hspost}@{}}%
\column{E}{@{}>{\hspre}l<{\hspost}@{}}%
\>[B]{}\mathbf{instance}\;\Conid{Monad}\;\Varid{m}\Rightarrow \Conid{Monad}\;(\Conid{CbV}\;\Varid{m})\;\mathbf{where}{}\<[E]%
\\
\>[B]{}\hsindent{3}{}\<[3]%
\>[3]{}\Varid{return}\;\Varid{v}\mathrel{=}\Conid{CbV}\;(\Varid{return}\;\Varid{v}){}\<[E]%
\\
\>[B]{}\hsindent{3}{}\<[3]%
\>[3]{}(\Conid{CbV}\;\Varid{a})\bind \Varid{f}\mathrel{=}\Conid{CbV}\;(\Varid{a}\bind (\Varid{runCbV}\mathbin{\circ}\Varid{f})){}\<[E]%
\\
\>[B]{}\mathbf{instance}\;\Conid{Monad}\;\Varid{m}\Rightarrow \Conid{Monad}\;(\Conid{CbN}\;\Varid{m})\;\mathbf{where}{}\<[E]%
\\
\>[B]{}\hsindent{3}{}\<[3]%
\>[3]{}\Varid{return}\;\Varid{v}\mathrel{=}\Conid{CbN}\;(\Varid{return}\;\Varid{v}){}\<[E]%
\\
\>[B]{}\hsindent{3}{}\<[3]%
\>[3]{}(\Conid{CbN}\;\Varid{a})\bind \Varid{f}\mathrel{=}\Conid{CbN}\;(\Varid{a}\bind (\Varid{runCbN}\mathbin{\circ}\Varid{f})){}\<[E]%
\ColumnHook
\end{hscode}\resethooks
\vspace{-2.2em}
\begin{hscode}\SaveRestoreHook
\column{B}{@{}>{\hspre}l<{\hspost}@{}}%
\column{E}{@{}>{\hspre}l<{\hspost}@{}}%
\>[B]{}\mathbf{instance}\;\Conid{MonadTrans}\;\Conid{CbV}\;\mathbf{where}\;\Varid{lift}\mathrel{=}\Conid{CbV}{}\<[E]%
\\
\>[B]{}\mathbf{instance}\;\Conid{MonadTrans}\;\Conid{CbN}\;\mathbf{where}\;\Varid{lift}\mathrel{=}\Conid{CbN}{}\<[E]%
\ColumnHook
\end{hscode}\resethooks
\vspace{-2.2em}
\begin{hscode}\SaveRestoreHook
\column{B}{@{}>{\hspre}l<{\hspost}@{}}%
\column{3}{@{}>{\hspre}l<{\hspost}@{}}%
\column{E}{@{}>{\hspre}l<{\hspost}@{}}%
\>[B]{}\mathbf{instance}\;\Conid{Monad}\;\Varid{m}\Rightarrow \Conid{MonadAlias}\;(\Conid{CbV}\;\Varid{m})\;\mathbf{where}{}\<[E]%
\\
\>[B]{}\hsindent{3}{}\<[3]%
\>[3]{}\Varid{malias}\;\Varid{m}\mathrel{=}\Varid{m}\bind (\Varid{return}\mathbin{\circ}\Varid{return}){}\<[E]%
\\
\>[B]{}\mathbf{instance}\;\Conid{Monad}\;\Varid{m}\Rightarrow \Conid{MonadAlias}\;(\Conid{CbN}\;\Varid{m})\;\mathbf{where}{}\<[E]%
\\
\>[B]{}\hsindent{3}{}\<[3]%
\>[3]{}\Varid{malias}\;\Varid{m}\mathrel{=}\Varid{return}\;\Varid{m}{}\<[E]%
\ColumnHook
\end{hscode}\resethooks
\vspace{-2em}
\caption{Instances of \ensuremath{\Conid{Monad}}, \ensuremath{\Conid{MonadTrans}} and \ensuremath{\Conid{MonadAlias}} for evaluation strategies}
\label{fig:monadalias-transformers}
\end{figure}

\paragraph{Example.} Using the previous definitions, we can now rewrite the example from
Section~\ref{sec:intro-translation} using generic functions that can be executed using both
\ensuremath{\Varid{runCbV}} and \ensuremath{\Varid{runCbN}}. Instead of implementing \ident{malias} for a specific monad such as \ensuremath{\Conid{IO}\;\Varid{a}}, 
we use a monad transformer \ensuremath{\Varid{t}} that lifts the monadic computation to either \ensuremath{\Conid{CbV}\;\Conid{IO}\;\Varid{a}} or to \ensuremath{\Conid{CbN}\;\Conid{IO}\;\Varid{a}}.
This means that all functions will have constraints \ensuremath{\Conid{MonadTrans}\;\Varid{t}}, specifying that \ensuremath{\Varid{t}}
is a monad transformer, and \ensuremath{\Conid{MonadAlias}\;(\Varid{t}\;\Varid{m})}, specifying that the computation implements
the \ident{malias} operation.

In Haskell, this can be succinctly written using \emph{constraint kinds} 
\cite{types-constraints-kinds}, that make it possible to define a single constraint
\ensuremath{\Conid{EvalStrategy}\;\Varid{t}\;\Varid{m}} that combines both of the conditions\footnote{Constraint kinds are
available in GHC 7.4 and generalize constraint families proposed by Orchard and
Schrijvers \cite{types-constraints-unleashed}}:

\begin{hscode}\SaveRestoreHook
\column{B}{@{}>{\hspre}l<{\hspost}@{}}%
\column{41}{@{}>{\hspre}l<{\hspost}@{}}%
\column{E}{@{}>{\hspre}l<{\hspost}@{}}%
\>[B]{}\mathbf{type}\;\Conid{EvalStrategy}\;\Varid{t}\;\Varid{m}\mathrel{=}(\Conid{MonadTrans}\;\Varid{t},{}\<[41]%
\>[41]{}\Conid{MonadAlias}\;(\Varid{t}\;\Varid{m})){}\<[E]%
\ColumnHook
\end{hscode}\resethooks
Despite the use of the \ensuremath{\mathbf{type}} keyword, the identifier \ensuremath{\Conid{EvalStrategy}} actually has a kind
\ensuremath{\Conid{Constraint}}, which means that it can be used to specify assumptions about types in a
function signature. In our example, the constraint \ensuremath{\Conid{EvalStrategy}\;\Varid{t}\;\Conid{IO}} denotes monadic
computations based on the \ensuremath{\Conid{IO}} monad that also provide an implementation of \ensuremath{\Conid{MonadAlias}}:

\begin{hscode}\SaveRestoreHook
\column{B}{@{}>{\hspre}l<{\hspost}@{}}%
\column{3}{@{}>{\hspre}l<{\hspost}@{}}%
\column{E}{@{}>{\hspre}l<{\hspost}@{}}%
\>[B]{}\Varid{chooseSize}\mathbin{::}\Conid{EvalStrategy}\;\Varid{t}\;\Conid{IO}\Rightarrow \Varid{t}\;\Conid{IO}\;\Conid{Int}\to \Varid{t}\;\Conid{IO}\;\Conid{Int}\to \Varid{t}\;\Conid{IO}\;\Conid{Int}{}\<[E]%
\\
\>[B]{}\Varid{chooseSize}\;\Varid{new}\;\Varid{legacy}\mathrel{=}\mathbf{do}{}\<[E]%
\\
\>[B]{}\hsindent{3}{}\<[3]%
\>[3]{}\Varid{newVal}\leftarrow \Varid{new}{}\<[E]%
\\
\>[B]{}\hsindent{3}{}\<[3]%
\>[3]{}\mathbf{if}\;\Varid{newVal}\mathbin{>}\mathrm{0}\;\mathbf{then}\;\Varid{new}\;\mathbf{else}\;\Varid{legacy}{}\<[E]%
\\[\blanklineskip]%
\>[B]{}\Varid{resultSize}\mathbin{::}\Conid{EvalStrategy}\;\Varid{t}\;\Conid{IO}\Rightarrow \Varid{t}\;\Conid{IO}\;\Conid{Int}{}\<[E]%
\\
\>[B]{}\Varid{resultSize}\mathrel{=}\mathbf{do}{}\<[E]%
\\
\>[B]{}\hsindent{3}{}\<[3]%
\>[3]{}\Varid{new}\leftarrow \Varid{malias}\mathbin{\$}\Varid{lift}\;(\Varid{lookupInput}\;\text{\tt \char34 new\char95 size\char34}){}\<[E]%
\\
\>[B]{}\hsindent{3}{}\<[3]%
\>[3]{}\Varid{legacy}\leftarrow \Varid{malias}\mathbin{\$}\Varid{lift}\;(\Varid{lookupInput}\;\text{\tt \char34 legacy\char95 size\char34}){}\<[E]%
\\
\>[B]{}\hsindent{3}{}\<[3]%
\>[3]{}\Varid{chooseSize}\;\Varid{new}\;\Varid{legacy}{}\<[E]%
\ColumnHook
\end{hscode}\resethooks
Compared to the previous version of the example, the only significant change is in the
type signature of the two functions. Instead of passing computations of type \ensuremath{\Conid{IO}\;\Varid{a}}, they
now work with computations \ensuremath{\Varid{t}\;\Conid{IO}\;\Varid{a}} with the constraint \ensuremath{\Conid{EvalStrategy}\;\Varid{t}\;\Conid{IO}}. The result of
\ensuremath{\Varid{lookupInput}} is still \ensuremath{\Conid{IO}\;\Conid{Int}} and so it needs to be lifted into a monadic computation 
\ensuremath{\Varid{t}\;\Conid{IO}\;\Conid{Int}} using the \ensuremath{\Varid{lift}} function.

The return type of the \ensuremath{\Varid{resultSize}} computation is parameterized over the evaluation strategy \ensuremath{\Varid{t}}.
This means that we can call it in two different ways. Writing \ensuremath{\Varid{runCbN}\;\Varid{resultSize}} executes the
function using \emph{call-by-name} and writing \ensuremath{\Varid{runCbV}\;\Varid{resultSize}} executes it using 
\emph{call-by-value}. However, it is also possible to implement a monad transformer 
for \emph{call-by-need}.

\subsection{Implementing call-by-need strategy}
\label{sec:practical-cbn}

In his paper introducing the \emph{call-by-name} and \emph{call-by-value} translations for 
monads, Wadler noted: ``It remains an open question whether there is a translation scheme 
that corresponds to call-by-need as opposed to call-by-name'' \cite{monads-wadler}.
We do not fully answer this question, however we hope to contribute to the answer.
In particular, we show how to use the mechanism described so far to turn certain monads
into an extended versions of such monads that provide the \emph{call-by-need} behaviour.

In the absence of effects, the \emph{call-by-need} strategy is equivalent to the 
\emph{call-by-name} strategy, with the only difference in that performance characteristics. 
In the \emph{call-by-need} (or \emph{lazy}) strategy, a computation passed as 
an argument is evaluated at most once and the result is cached afterwards. 

The caching of results needs to be done in the \ident{malias} operation. This cannot be 
done for \emph{any} monad, but we can define a monad transformer similar to the ones 
presented in the previous section. In particular, we use Svenningsson's package 
\cite{monad-sttrans-package}, which defines a transformer version of the \ensuremath{\Conid{ST}} monad \cite{monad-state-haskell}.
As documented in the package description, the transformer should be applied only to monads
that yield a single result. Combining lazy evaluation with non-determinism is a more 
complex topic that has been explored by Fischer et\ al. \cite{monad-pure-nondeterministic}.

\begin{hscode}\SaveRestoreHook
\column{B}{@{}>{\hspre}l<{\hspost}@{}}%
\column{E}{@{}>{\hspre}l<{\hspost}@{}}%
\>[B]{}\mathbf{newtype}\;\Conid{CbL}\;\Varid{s}\;\Varid{m}\;\Varid{a}\mathrel{=}\Conid{CbL}\;\{\mskip1.5mu \Varid{unCbL}\mathbin{::}\Conid{STT}\;\Varid{s}\;\Varid{m}\;\Varid{a}\mskip1.5mu\}{}\<[E]%
\ColumnHook
\end{hscode}\resethooks
Unlike \ensuremath{\Conid{CbV}} and \ensuremath{\Conid{CbN}}, the \ensuremath{\Conid{CbL}} type is not a simple wrapper that contains a computation
of type \ensuremath{\Varid{m}\;\Varid{a}}. Instead, it contains a computation augmented with some additional state.
The state is used for caching the values of evaluated computations. The type \ensuremath{\Conid{STT}\;\Varid{s}\;\Varid{m}\;\Varid{a}}
represents a computation \ensuremath{\Varid{m}\;\Varid{a}} with an additional local state ``tagged'' with a type variable \ensuremath{\Varid{s}}.
The use of a local state instead of e.g.\ \ensuremath{\Conid{IO}} means that the monadic computation can be 
safely evaluated even as part of purely functional code. The tags are used merely to guarantee 
that state associated with one \ensuremath{\Conid{STT}} computation does not leak to other parts of the program.

Implementing the \ensuremath{\Conid{Monad}} and \ensuremath{\Conid{MonadTrans}} instances follows exactly the same pattern as 
instances of other transformers given in Figure~\ref{fig:monadalias-transformers}.
The interesting work is done in the \ident{malias} function of \ensuremath{\Conid{MonadAlias}}:

\begin{hscode}\SaveRestoreHook
\column{B}{@{}>{\hspre}l<{\hspost}@{}}%
\column{3}{@{}>{\hspre}l<{\hspost}@{}}%
\column{5}{@{}>{\hspre}l<{\hspost}@{}}%
\column{7}{@{}>{\hspre}l<{\hspost}@{}}%
\column{9}{@{}>{\hspre}l<{\hspost}@{}}%
\column{E}{@{}>{\hspre}l<{\hspost}@{}}%
\>[B]{}\mathbf{instance}\;\Conid{Monad}\;\Varid{m}\Rightarrow \Conid{MonadAlias}\;(\Conid{CbL}\;\Varid{s}\;\Varid{m})\;\mathbf{where}{}\<[E]%
\\
\>[B]{}\hsindent{3}{}\<[3]%
\>[3]{}\Varid{malias}\;(\Conid{CbL}\;\Varid{marg})\mathrel{=}\Conid{CbL}\mathbin{\$}\mathbf{do}{}\<[E]%
\\
\>[3]{}\hsindent{2}{}\<[5]%
\>[5]{}\Varid{r}\leftarrow \Varid{newSTRef}\;\Conid{Nothing}{}\<[E]%
\\
\>[3]{}\hsindent{2}{}\<[5]%
\>[5]{}\Varid{return}\;(\Conid{CbL}\mathbin{\$}\mathbf{do}{}\<[E]%
\\
\>[5]{}\hsindent{2}{}\<[7]%
\>[7]{}\Varid{rv}\leftarrow \Varid{readSTRef}\;\Varid{r}{}\<[E]%
\\
\>[5]{}\hsindent{2}{}\<[7]%
\>[7]{}\mathbf{case}\;\Varid{rv}\;\mathbf{of}{}\<[E]%
\\
\>[7]{}\hsindent{2}{}\<[9]%
\>[9]{}\Conid{Nothing}\to \Varid{marg}\bind \lambda \Varid{v}\to \Varid{writeSTRef}\;\Varid{r}\;(\Conid{Just}\;\Varid{v})\sequ \Varid{return}\;\Varid{v}{}\<[E]%
\\
\>[7]{}\hsindent{2}{}\<[9]%
\>[9]{}\Conid{Just}\;\Varid{v}\to \Varid{return}\;\Varid{v}){}\<[E]%
\ColumnHook
\end{hscode}\resethooks
The \ident{malias} operation takes a computation of type \ensuremath{\Varid{m}\;\Varid{a}} and returns a computation
\ensuremath{\Varid{m}\;(\Varid{m}\;\Varid{a})}. The monad transformer wraps the underlying monad inside \ensuremath{\Conid{STT}}, so the type of the
computation returned by \ensuremath{\Varid{malias}} is equivalent to a type \ensuremath{\Conid{STT}\;\Varid{s}\;\Varid{m}\;(\Conid{STT}\;\Varid{s}\;\Varid{m}\;\Varid{a})}. 

The fact that both outer and inner \ensuremath{\Conid{STT}} 
share the same tag \ensuremath{\Varid{s}} means that they operate on a shared state. The outer computation
allocates a new reference and the inner computation can use it to access and store the
result computed previously. The allocation is done using the \ensuremath{\Varid{newSTRef}} function, which 
creates a reference initialized to \ensuremath{\Conid{Nothing}}. In the returned (inner) computation, we 
first read the state using \ensuremath{\Varid{readSTRef}}. If the value was computed previously, then it is
simply returned. If not, the computation evaluates \ensuremath{\Varid{marg}}, stores the result in a reference
cell and then returns the obtained value.

\paragraph{Discussion.}
After implementing \ensuremath{\Varid{runCbL}} function (which can be done easily using \ensuremath{\Varid{runST}}), the example
in Section~\ref{sec:practical-strategypolymorphic} can be executed using the \ensuremath{\Conid{CbL}} type.
With the \emph{call-by-need} semantics, the program finally behaves as desired:
if the value of the \ensuremath{\text{\tt \char34 new\char95 size\char34}} key is greater than zero, then it reads it only 
once, without reading the value of the \ensuremath{\text{\tt \char34 legacy\char95 size\char34}} key. The value of \ensuremath{\text{\tt \char34 legacy\char95 size\char34}}
key is accessed only if the value of the \ensuremath{\text{\tt \char34 new\char95 size\char34}} key is less than zero.

Showing that the above definition corresponds to \emph{call-by-need} formally is beyond
the scope of this paper. However, the use of \ensuremath{\Conid{STT}} transformer adds a shared state that
keeps the evaluated values, which closely corresponds to Launchbury's environment-based 
semantics of lazy evaluation \cite{formalism-lazy-semantics}. 

We do not formally prove that the \ensuremath{\Varid{malias}} laws hold for the above implementation, but we 
give an informal argument. The \emph{naturality} law follows from parametricity. 
Computations considered in the \emph{associativity} law are of type \ensuremath{\Varid{m}\;(\Varid{m}\;(\Varid{m}\;\Varid{a}))}; both
sides of the equation create a computation where the two outer \ensuremath{\Varid{m}} computations allocate
a new reference (where the outer points to the inner) and the single innermost \ensuremath{\Varid{m}} actually
triggers the computation. The \emph{computationality} law holds, because aliasing of a
unit computation cannot introduce any effects. Finally, the left-hand side of 
\emph{identity} creates a computation that allocates a new reference that is encapsulated
in the returned computation and cannot be accessed from elsewhere, so no sharing is added 
when compared with the right-hand side.

\subsection{Parallel call-by-need strategy}
\label{sec:practical-parallel-cbn}

In this section, we consider yet another evaluation strategy that can be implemented using
our scheme. The \emph{parallel call-by-need} strategy \cite{parallel-cbn-semantics}
is similar to \emph{call-by-need}, but it may optimistically start evaluating a computation 
sooner, in parallel with the main program. When carefully tuned, the evaluation strategy
may result in a better performance on multi-core CPUs.

We present a simple implementation of the \ident{malias} operation based on the monad for
deterministic parallelism by Marlow et al. \cite{monad-parallel-det}. By translating purely 
functional code to a monadic version using our translation and the \ensuremath{\Conid{Par}} monad, we get a 
program that attempts to evaluate arguments of every function call in parallel. In practice, 
this may introduce too much overhead, but it demonstrate that \emph{parallel call-by-need}
strategy also fits with our general framework.

Unlike the previous two sections, we do not define a monad transformer that can embody
any monadic computation. For example, performing IO operations in parallel might introduce
non-determinism. Instead, we implement \ident{malias} operation directly for computations
of type \ensuremath{\Conid{Par}\;\Varid{a}}:

\begin{hscode}\SaveRestoreHook
\column{B}{@{}>{\hspre}l<{\hspost}@{}}%
\column{3}{@{}>{\hspre}l<{\hspost}@{}}%
\column{E}{@{}>{\hspre}l<{\hspost}@{}}%
\>[B]{}\mathbf{instance}\;\Conid{MonadAlias}\;\Conid{Par}\;\mathbf{where}{}\<[E]%
\\
\>[B]{}\hsindent{3}{}\<[3]%
\>[3]{}\Varid{malias}\;\Varid{m}\mathrel{=}\Varid{spawn}\;\Varid{m}\bind \Varid{return}\mathbin{\circ}\Varid{get}{}\<[E]%
\ColumnHook
\end{hscode}\resethooks
The implementation is surprisingly simple. The function \ensuremath{\Varid{spawn}} creates a computation \ensuremath{\Conid{Par}\;(\Conid{IVar}\;\Varid{a})}
that starts the work in background and returns a mutable variable (I-structure) that will
contain the result when the computation completes. The inner computation that is returned by
\ident{malias} calls a function \ensuremath{\Varid{get}\mathbin{::}\Conid{IVar}\;\Varid{a}\to \Conid{Par}\;\Varid{a}} that waits until \ensuremath{\Conid{IVar}} has been assigned
a value and then returns it.

Using the above implementation of \ensuremath{\Varid{malias}}, we can now translate purely functional code to a 
monadic version that uses \emph{parallel call-by-need} instead of the previous standard evaluation 
strategies (that do not introduce any parallelism). For example, consider a naive Fibonacci function:

\begin{hscode}\SaveRestoreHook
\column{B}{@{}>{\hspre}l<{\hspost}@{}}%
\column{E}{@{}>{\hspre}l<{\hspost}@{}}%
\>[B]{}\Varid{fibSeq}\;\Varid{n}\mid \Varid{n}\leq \mathrm{1}\mathrel{=}\Varid{n}{}\<[E]%
\\
\>[B]{}\Varid{fibSeq}\;\Varid{n}\mathrel{=}\Varid{fibSeq}\;(\Varid{n}\mathbin{-}\mathrm{1})\mathbin{+}\Varid{fibSeq}\;(\Varid{n}\mathbin{-}\mathrm{2}){}\<[E]%
\ColumnHook
\end{hscode}\resethooks
The second case calls the \ensuremath{\mathbin{+}} operator with two computations as arguments. The translated version
passes these computations to \ident{malias}, which starts executing them in parallel. The monadic
version of the \ensuremath{\mathbin{+}} operator then waits until both computations complete. If we translate only 
the second case and manually add a case that calls sequential version of the function for inputs
smaller than 30, we get the following:

\begin{hscode}\SaveRestoreHook
\column{B}{@{}>{\hspre}l<{\hspost}@{}}%
\column{5}{@{}>{\hspre}l<{\hspost}@{}}%
\column{E}{@{}>{\hspre}l<{\hspost}@{}}%
\>[B]{}\Varid{fibPar}\;\Varid{n}\mid \Varid{n}\mathbin{<}\mathrm{30}\mathrel{=}\Varid{return}\;(\Varid{fibSeq}\;\Varid{n}){}\<[E]%
\\
\>[B]{}\Varid{fibPar}\;\Varid{n}\mathrel{=}\mathbf{do}{}\<[E]%
\\
\>[B]{}\hsindent{5}{}\<[5]%
\>[5]{}\Varid{n1}\leftarrow \Varid{malias}\mathbin{\$}\Varid{fibPar}\;(\Varid{n}\mathbin{-}\mathrm{1}){}\<[E]%
\\
\>[B]{}\hsindent{5}{}\<[5]%
\>[5]{}\Varid{n2}\leftarrow \Varid{malias}\mathbin{\$}\Varid{fibPar}\;(\Varid{n}\mathbin{-}\mathrm{2}){}\<[E]%
\\
\>[B]{}\hsindent{5}{}\<[5]%
\>[5]{}\Varid{liftM2}\;(\mathbin{+})\;\Varid{n1}\;\Varid{n2}{}\<[E]%
\ColumnHook
\end{hscode}\resethooks
Aside from the first line, the code directly follows the general translation mechanism described earlier.
Arguments of a function are turned to monadic computations and passed to \ident{malias}. The
inner computations obtained using monadic bind are then passed to a translated function. 

Thanks to the manual optimization that calls \ensuremath{\Varid{fibSeq}} for smaller inputs, the function
runs nearly twice as fast on a dual-core machine\footnote{When called with input 37 on a
Core 2 Duo CPU, the sequential version runs in 8.9s and the parallel version in 5.1s.}. 
Parallel programming is one of the first areas where we found the \ident{malias} operation 
useful. We first considered it as part of \emph{joinads}, which are discussed in the next section.

\paragraph{Discussion.}
Showing that the above implementation actually implements \emph{parallel call-by-need} could be done
by relating our implementation of \ensuremath{\Varid{malias}} to the multi-thread transitions of \cite{parallel-cbn-semantics}.
Informally, each computation that may be shared is added to the work queue (using \ensuremath{\Varid{spawn}}) when it 
occurs on the right-hand side of a let binding, or as an argument in function application. The 
queued work evaluates in parallel with the main program and the \ensuremath{\Varid{get}} function implements sharing,
so the semantics is \emph{lazy}.

As previously, the \emph{naturality} law holds thanks to parametricity.
To consider other laws, we need a formal model that captures the time needed to evaluate computations. 
We assume that evaluating a computation created by \ensuremath{\Varid{unit}} takes no time, but all other 
computations take non-zero time. Moreover, all spawned computations start executing immediately (i.e.\ the 
number of equally fast threads is unlimited).

In the \emph{associativity} law,
the left-hand side returns a computation that spawns the actual work and then spawns a computation
that waits for its completion. The right-hand side returns a computation that schedules a computation, 
which then schedules the actual work. In both cases, the actual work is started immediately when the
outer \ensuremath{\Varid{m}} computation is evaluated, and so they are equivalent.
The \emph{computationality} law holds, because a \ensuremath{\Varid{unit}} computation evaluates in no time. Finally, the
left-hand side of \emph {identity} returns a computation that spawns the work and then waits for its 
completion, which is semantically equivalent to just running the computation.

\subsection{Simplifying Joinads}
\label{sec:practical-joinads}
Joinads \cite{joinads,joinads-haskell11} were designed to simplify programming with certain kinds
of monadic computations. Many monads, especially from the area of concurrent or parallel programming
provide additional operations for composing monadic computations. Joinads identify three most common
extensions: \emph{parallel composition}, \emph{(non-deterministic) choice} and \emph{aliasing}. 

Joinads are abstract computations that form a monad and provide the three additional operations
mentioned above. The work on joinads also introduces a syntactic extension for Haskell and F\# 
that makes it easier to work with these classes of computations. For example, the following 
snippet uses the \ensuremath{\Conid{Par}} monad to implement a function that tests whether a predicate holds for 
all leafs of a tree in parallel:

\begin{hscode}\SaveRestoreHook
\column{B}{@{}>{\hspre}l<{\hspost}@{}}%
\column{3}{@{}>{\hspre}l<{\hspost}@{}}%
\column{5}{@{}>{\hspre}l<{\hspost}@{}}%
\column{19}{@{}>{\hspre}l<{\hspost}@{}}%
\column{26}{@{}>{\hspre}l<{\hspost}@{}}%
\column{E}{@{}>{\hspre}l<{\hspost}@{}}%
\>[B]{}\Varid{all}\mathbin{::}(\Varid{a}\to \Conid{Bool})\to \Conid{Tree}\;\Varid{a}\to \Conid{Par}\;\Conid{Bool}{}\<[E]%
\\[\blanklineskip]%
\>[B]{}\Varid{all}\;\Varid{p}\;(\Conid{Leaf}\;\Varid{v}){}\<[26]%
\>[26]{}\mathrel{=}\Varid{return}\;(\Varid{p}\;\Varid{v}){}\<[E]%
\\
\>[B]{}\Varid{all}\;\Varid{p}\;(\Conid{Node}\;\Varid{left}\;\Varid{right}){}\<[26]%
\>[26]{}\mathrel{=}{}\<[E]%
\\
\>[B]{}\hsindent{3}{}\<[3]%
\>[3]{}\mathbf{docase}\;(\Varid{all}\;\Varid{p}\;\Varid{left},\Varid{all}\;\Varid{p}\;\Varid{right})\;\mathbf{of}{}\<[E]%
\\
\>[3]{}\hsindent{2}{}\<[5]%
\>[5]{}(\Conid{False},\mathbin{?}){}\<[19]%
\>[19]{}\to \Varid{return}\;\Conid{False}{}\<[E]%
\\
\>[3]{}\hsindent{2}{}\<[5]%
\>[5]{}(\mathbin{?},\Conid{False}){}\<[19]%
\>[19]{}\to \Varid{return}\;\Conid{False}{}\<[E]%
\\
\>[3]{}\hsindent{2}{}\<[5]%
\>[5]{}(\Varid{allL},\Varid{allR}){}\<[19]%
\>[19]{}\to \Varid{return}\;(\Varid{allL}\mathrel{\wedge}\Varid{allR}){}\<[E]%
\ColumnHook
\end{hscode}\resethooks
The \ensuremath{\mathbf{docase}} notation intentionally resembles pattern matching and has similar semantics as well.
The first two cases use the special pattern \ensuremath{\mathbin{?}} to denote that the value of one of the computations does
not have to be available in order to continue. When one of the sub-branches returns \ensuremath{\Conid{False}}, we know
that the overall result is \ensuremath{\Conid{False}} and so we return immediately. Finally, the last clause matches
if neither of the two previous are matched. It can only match after both sub-trees are processed.

Similarly to \ensuremath{\mathbf{do}} notation, the \ensuremath{\mathbf{docase}} syntax is desugared into uses of the joinad operations.
The choice between clauses is translated using the \emph{choice} operator. If a clause requires the
result of multiple computations (such as the last one), the computations are combined using 
\emph{parallel composition}.

If a computation, passed as an argument, is accessed from multiple clauses, then it should be 
evaluated only once and the clauses should only access \emph{aliased} computation. This motivation
is similar to the one described in this article. Indeed, joinads use a variant of the \ident{malias}
operation and insert it automatically for all arguments of \ensuremath{\mathbf{docase}}. This is very similar to how
the translation presented in this paper uses \ident{malias}. If aliasing was not done automatically
behind the scenes, we would have to write:

\begin{hscode}\SaveRestoreHook
\column{B}{@{}>{\hspre}l<{\hspost}@{}}%
\column{3}{@{}>{\hspre}l<{\hspost}@{}}%
\column{5}{@{}>{\hspre}l<{\hspost}@{}}%
\column{19}{@{}>{\hspre}l<{\hspost}@{}}%
\column{26}{@{}>{\hspre}l<{\hspost}@{}}%
\column{E}{@{}>{\hspre}l<{\hspost}@{}}%
\>[B]{}\Varid{all}\;\Varid{p}\;(\Conid{Leaf}\;\Varid{v}){}\<[26]%
\>[26]{}\mathrel{=}\Varid{return}\;(\Varid{p}\;\Varid{v}){}\<[E]%
\\
\>[B]{}\Varid{all}\;\Varid{p}\;(\Conid{Node}\;\Varid{left}\;\Varid{right}){}\<[26]%
\>[26]{}\mathrel{=}\mathbf{do}{}\<[E]%
\\
\>[B]{}\hsindent{3}{}\<[3]%
\>[3]{}\Varid{l}\leftarrow \Varid{malias}\;(\Varid{all}\;\Varid{p}\;\Varid{left}){}\<[E]%
\\
\>[B]{}\hsindent{3}{}\<[3]%
\>[3]{}\Varid{r}\leftarrow \Varid{malias}\;(\Varid{all}\;\Varid{p}\;\Varid{right}){}\<[E]%
\\
\>[B]{}\hsindent{3}{}\<[3]%
\>[3]{}\mathbf{docase}\;(\Varid{l},\Varid{r})\;\mathbf{of}{}\<[E]%
\\
\>[3]{}\hsindent{2}{}\<[5]%
\>[5]{}(\Conid{False},\mathbin{?}){}\<[19]%
\>[19]{}\to \Varid{return}\;\Conid{False}{}\<[E]%
\\
\>[3]{}\hsindent{2}{}\<[5]%
\>[5]{}(\mathbin{?},\Conid{False}){}\<[19]%
\>[19]{}\to \Varid{return}\;\Conid{False}{}\<[E]%
\\
\>[3]{}\hsindent{2}{}\<[5]%
\>[5]{}(\Varid{allL},\Varid{allR}){}\<[19]%
\>[19]{}\to \Varid{return}\;(\Varid{allL}\mathrel{\wedge}\Varid{allR}){}\<[E]%
\ColumnHook
\end{hscode}\resethooks
One of the limitations of the original design of joinads is that there is no one-to-one correspondence
between the \ensuremath{\mathbf{docase}} notation and what can be expressed directly using the joinad operations.
This is partly due to the automatic aliasing of arguments, which inserts \ident{malias} only at 
very specific locations. We believe that integrating a \emph{call-by-alias} translation, described
in this article, in a programming language could resolve this situation. It would also separate the 
two concerns -- composition of computations using \emph{choice} and \emph{parallel composition} 
(done by joinads) and automatic aliasing of computations that allows sharing of results as in 
\emph{call-by-need} or \emph{parallel call-by-need}.

\section{Related work}
\label{sec:related-work}

Most of the work that directly influenced this work has been discussed throughout the 
paper. Most importantly, the question of translating pure code to a monadic version with
\emph{call-by-need} semantics was posed by Wadler \cite{monads-wadler}. To our knowledge,
this question has not been answered before, but there is various work that 
either uses similar structures or considers evaluation strategies from 
different perspectives.

\paragraph{Monads with aliasing.}
Numerous monads have independently introduced an operation of type \ensuremath{\Varid{m}\;\Varid{a}\to \Varid{m}\;(\Varid{m}\;\Varid{a})}.
The \ensuremath{\Varid{eagerly}} combinator of the Orc monad \cite{monad-orc} causes computation to run in 
parallel, effectively implementing the \emph{parallel call-by-need} evaluation strategy. The
only law that relates \ensuremath{\Varid{eagerly}} to operations of a monad is too restrictive to allow 
the \emph{call-by-value} semantics.

A monad for purely functional lazy non-deterministic programming \cite{monad-pure-nondeterministic}
uses a similar combinator \ensuremath{\Varid{share}} to make monadic (non-deterministic) computations
lazy. However, the \emph{call-by-value} strategy is inefficient and the \emph{call-by-name}
strategy is incorrect, because choosing a different value each time a non-deterministic
computation is accessed means that \emph{generate and test} pattern does not work.

The \ensuremath{\Varid{share}} operation is described together with the laws that should hold. The \emph{HNF}
law is similar to our \emph{computationality}. However, the \emph{Ignore} law specifies
that the \ensuremath{\Varid{share}} operation should not be strict (ruling out our \emph{call-by-value} implementation
of \ensuremath{\Varid{malias}}). A related paper \cite{monad-pure-nondeterministic-translating}
discusses where \ensuremath{\Varid{share}} needs to be inserted when translating lazy non-deterministic programs
to a monadic form. The results may be directly applicable to make our translation
more efficient by inserting \ensuremath{\Varid{malias}} only when required.

\paragraph{Abstract computations and comonads.}
In this paper, we extended monads with one component of a comonadic structure. Although less 
widespread than monads, comonads are also useful for capturing abstract computations in 
functional programming. They have been used for dataflow programming \cite{comonads-dataflow-essence}, 
array programming \cite{comonads-ypnos}, environment passing, and more \cite{comonads-codata-haskell}.
In general, comonads can be used to describe context-dependent computations \cite{comonads-notions},
where \ensuremath{\Varid{cojoin}} (natural transformation $\delta$) duplicates the context. In our work, the
corresponding operation \ident{malias} splits the context (effects) in a particular way
between two computations.

We only considered basic monadic computations, but it would be interesting to see how
\ident{malias} interacts with other abstract notions of computations, such as 
\emph{applicative functors} \cite{applicative-programming}, \emph{arrows} \cite{arrows-generalising-monads}
or additive monads (the \ensuremath{\Conid{MonadPlus}} type-class). The monad for lazy non-deterministic
programming \cite{monad-pure-nondeterministic}, mentioned earlier, implements 
\ensuremath{\Conid{MonadPlus}} and may thus provide interesting insights.

\paragraph{Evaluation strategies.}  
One of the key results of this paper is a monadic translation from purely 
functional code to a monadic version that has the \emph{call-by-need} semantics. We achieve
that using the monad transformer \cite{monad-transformer-interpreters} for adding state.

In the absence of effects, \emph{call-by-need} is equivalent to \emph{call-by-name}, but it has been
described formally as a version of $\lambda$-calculus by Ariola and Felleisen 
\cite{formalism-callbyneed-calculus}. This allows equational reasoning about computations
and it could be used to show that our encoding directly corresponds to \emph{call-by-need},
similarly to proofs for other strategies in Appendix~\ref{sec:appendix-cbv-cbn-equivalence}.
The semantics has been also described using an environment that models caching
\cite{formalism-lazy-semantics,formalism-lazy-sharing-semantics}, which closely corresponds
to our actual implementation.

Considering the two basic evaluation strategies, Wadler \cite{formalism-cbn-dual-cbv} shows that 
\emph{call-by-name} is dual to \emph{call-by-value}. We find this curious as the two
definitions of \ident{malias} in our work are, in some sense, also dual or symmetric as they 
associate all effects with the inner or the outer monad of the type \ensuremath{\Varid{m}\;(\Varid{m}\;\Varid{a})}. Furthermore, the
duality between \emph{call-by-name} and \emph{call-by-value} can be viewed from a logical 
perspective thanks to the Curry-Howard correspondence. We believe that finding a similar logical
perspective for our generalized strategy may be an interesting future work.

Finally, the work presented in this work unifies monadic \emph{call-by-name} and 
\emph{call-by-value}. In a non-monadic setting, a similar goal is achieved by the
\emph{call-by-push-value} calculus \cite{formalism-cbpv}. The calculus is more fine-grained 
and strictly separates \emph{values} and \emph{computations}. Using these mechanisms,
it is possible to encode both \emph{call-by-name} and \emph{call-by-value}. It may be
interesting to consider whether our computations parameterized over evaluation strategy
(Section~\ref{sec:practical-strategypolymorphic}) could be encoded in \emph{call-by-push-value}.


\section{Conclusions}

We presented an alternative translation from purely functional code to 
monadic form. Previously, this required choosing either the \emph{call-by-need} or the 
\emph{call-by-value} translation and the translated code had different structure and 
different types in both cases. Our translation abstracts the evaluation strategy into
a function \ident{malias} that can be implemented separately providing the required
evaluation strategy.

Our translation is not limited to the above two evaluation strategies. Most interestingly,
we show that certain monads can be automatically turned into an extended version that
supports the \emph{call-by-need} strategy. This answers part of an interesting open
problem posed by Wadler \cite{monads-wadler}. The approach has other
interesting applications -- it makes it possible to write code that is parameterized by
the evaluation strategy and it allows implementing a \emph{parallel call-by-need} strategy
for certain monads.

Finally, we presented the theoretical foundations of our approach using a model 
described in terms of category theory. We extended the monad structure with an additional operation
based on \emph{computational comonads}, which were previously used to give intensional semantics
of computations. In our setting, the operation specifies the evaluation order. 
The categorical model specifies laws about \ident{malias} and we proved that the laws hold 
for \emph{call-by-value} and \emph{call-by-name} strategies.

\paragraph{Acknowledgements.} The author is grateful to Sebastian Fischer, Dominic Orchard
and Alan Mycroft for inspiring comments and discussion, and to the latter two for proofreading 
the paper. We are also grateful to anonymous referees for detailed comments and to
Paul Blain Levy for shepherding of the paper. The work has been partly supported by EPSRC 
and through the Cambridge CHESS scheme.


\bibliographystyle{eptcs}
\bibliography{malias}

\begin{thebibliography}{10}
\providecommand{\bibitemdeclare}[2]{}
\providecommand{\urlprefix}{Available at }
\providecommand{\url}[1]{\texttt{#1}}
\providecommand{\href}[2]{\texttt{#2}}
\providecommand{\urlalt}[2]{\href{#1}{#2}}
\providecommand{\doi}[1]{doi:\urlalt{http://dx.doi.org/#1}{#1}}
\providecommand{\bibinfo}[2]{#2}

\bibitemdeclare{article}{formalism-callbyneed-calculus}
\bibitem{formalism-callbyneed-calculus}
\bibinfo{author}{Zena~M. Ariola} \& \bibinfo{author}{Matthias Felleisen}
  (\bibinfo{year}{1997}): \emph{\bibinfo{title}{The {C}all-{B}y-{N}eed {L}ambda
  {C}alculus}}.
\newblock {\sl \bibinfo{journal}{Journal of Functional Programming}}
  \bibinfo{volume}{7}, pp. \bibinfo{pages}{265--301},
  \doi{10.1017/S0956796897002724}.

\bibitemdeclare{inproceedings}{parallel-cbn-semantics}
\bibitem{parallel-cbn-semantics}
\bibinfo{author}{Clem Baker-Finch}, \bibinfo{author}{David~J. King} \&
  \bibinfo{author}{Phil Trinder} (\bibinfo{year}{2000}):
  \emph{\bibinfo{title}{An {O}perational {S}emantics for {P}arallel {L}azy
  {E}valuation}}.
\newblock In: {\sl \bibinfo{booktitle}{Proceedings of ICFP 2000}},
  \doi{10.1145/351240.351256}.

\bibitemdeclare{unpublished}{types-constraints-kinds}
\bibitem{types-constraints-kinds}
\bibinfo{author}{Max Bolingbroke} (\bibinfo{year}{2011}):
  \emph{\bibinfo{title}{Constraint Kinds for GHC (Unpublished manuscript)}}.
\newblock \urlprefix\url{http://blog.omega-prime.co.uk/?p=127}.

\bibitemdeclare{inproceedings}{monad-pure-nondeterministic-translating}
\bibitem{monad-pure-nondeterministic-translating}
\bibinfo{author}{B.~Bra{\ss}el}, \bibinfo{author}{S.~Fischer},
  \bibinfo{author}{M.~Hanus} \& \bibinfo{author}{F.~Reck}
  (\bibinfo{year}{2011}): \emph{\bibinfo{title}{Transforming Functional Logic
  Programs into Monadic Functional Programs}}.
\newblock In: {\sl \bibinfo{booktitle}{Proceedings of WFLP 2010}},
  \bibinfo{publisher}{LNCS 6559}, pp. \bibinfo{pages}{30--47},
  \doi{10.1.1.157.4578}.

\bibitemdeclare{inproceedings}{comonads-computational}
\bibitem{comonads-computational}
\bibinfo{author}{Stephen Brookes} \& \bibinfo{author}{Shai Geva}
  (\bibinfo{year}{1992}): \emph{\bibinfo{title}{{Computational Comonads and
  Intensional Semantics}}}.
\newblock In: {\sl \bibinfo{booktitle}{Applications of Categories in Computer
  Science: Proceedings of the LMS Symposium}}, \bibinfo{volume}{177}, pp.
  \bibinfo{pages}{1--44}, \doi{10.1.1.45.4952}.

\bibitemdeclare{article}{monad-pure-nondeterministic}
\bibitem{monad-pure-nondeterministic}
\bibinfo{author}{Sebastian Fischer}, \bibinfo{author}{Oleg Kiselyov} \&
  \bibinfo{author}{Chung-chieh Shan} (\bibinfo{year}{2011}):
  \emph{\bibinfo{title}{{Purely Functional Lazy Non-Deterministic
  Programming}}}.
\newblock {\sl \bibinfo{journal}{Journal of Functional Programming}}
  \bibinfo{volume}{21}(\bibinfo{number}{4--5}), pp. \bibinfo{pages}{413--465},
  \doi{10.1145/1631687.1596556}.

\bibitemdeclare{article}{arrows-generalising-monads}
\bibitem{arrows-generalising-monads}
\bibinfo{author}{John Hughes} (\bibinfo{year}{1998}):
  \emph{\bibinfo{title}{Generalising Monads to Arrows}}.
\newblock {\sl \bibinfo{journal}{Science of Computer Programming}}
  \bibinfo{volume}{37}, pp. \bibinfo{pages}{67--111},
  \doi{10.1016/S0167-6423(99)00023-4}.

\bibitemdeclare{unpublished}{comonads-codata-haskell}
\bibitem{comonads-codata-haskell}
\bibinfo{author}{Richard~B. Kieburtz} (\bibinfo{year}{1999}):
  \emph{\bibinfo{title}{Codata and Comonads in Haskell (Unpublished
  manuscript)}}.

\bibitemdeclare{inproceedings}{formalism-lazy-semantics}
\bibitem{formalism-lazy-semantics}
\bibinfo{author}{John Launchbury} (\bibinfo{year}{1993}):
  \emph{\bibinfo{title}{{A Natural Semantics for Lazy Evaluation}}}.
\newblock In: {\sl \bibinfo{booktitle}{Proceedings of POPL 1993}}, pp.
  \bibinfo{pages}{144--154}, \doi{10.1145/158511.158618}.

\bibitemdeclare{inproceedings}{monad-orc}
\bibitem{monad-orc}
\bibinfo{author}{John Launchbury} \& \bibinfo{author}{Trevor Elliott}
  (\bibinfo{year}{2010}): \emph{\bibinfo{title}{Concurrent orchestration in
  Haskell}}.
\newblock In: {\sl \bibinfo{booktitle}{Proceedings of Haskell Symposium}},
  \bibinfo{series}{Haskell 2010}, pp. \bibinfo{pages}{79--90},
  \doi{10.1145/1863523.1863534}.

\bibitemdeclare{inproceedings}{monad-state-haskell}
\bibitem{monad-state-haskell}
\bibinfo{author}{John Launchbury} \& \bibinfo{author}{Simon~Peyton Jones}
  (\bibinfo{year}{1995}): \emph{\bibinfo{title}{State in Haskell}}.
\newblock In: {\sl \bibinfo{booktitle}{Proceedings of the LISP and Symbolic
  Computation Conference}}, \bibinfo{series}{LISP 1995},
  \doi{10.1007/BF01018827}.

\bibitemdeclare{book}{formalism-cbpv}
\bibitem{formalism-cbpv}
\bibinfo{author}{Paul~Blain Levy} (\bibinfo{year}{2004}):
  \emph{\bibinfo{title}{Call-By-Push-Value}}.
\newblock \bibinfo{publisher}{Springer}.
\newblock \bibinfo{note}{ISBN: 978-1-4020-1730-8}.

\bibitemdeclare{inproceedings}{monad-transformer-interpreters}
\bibitem{monad-transformer-interpreters}
\bibinfo{author}{Sheng Liang}, \bibinfo{author}{Paul Hudak} \&
  \bibinfo{author}{Mark Jones} (\bibinfo{year}{1995}):
  \emph{\bibinfo{title}{{Monad Transformers and Modular Interpreters}}}.
\newblock In: {\sl \bibinfo{booktitle}{Proceedings of POPL 1995}},
  \doi{10.1145/199448.199528}.

\bibitemdeclare{inproceedings}{monad-parallel-det}
\bibitem{monad-parallel-det}
\bibinfo{author}{Simon Marlow}, \bibinfo{author}{Ryan Newton} \&
  \bibinfo{author}{Simon Peyton~Jones} (\bibinfo{year}{2011}):
  \emph{\bibinfo{title}{{A Monad for Deterministic Parallelism}}}.
\newblock In: {\sl \bibinfo{booktitle}{Proceedings of the 4th Symposium on
  Haskell}}, \bibinfo{series}{Haskell 2011}, \doi{10.1145/2034675.2034685}.

\bibitemdeclare{article}{applicative-programming}
\bibitem{applicative-programming}
\bibinfo{author}{Conor Mc{B}ride} \& \bibinfo{author}{Ross Paterson}
  (\bibinfo{year}{2008}): \emph{\bibinfo{title}{{Applicative Programming with
  Effects}}}.
\newblock {\sl \bibinfo{journal}{Journal of Functional Programming}}
  \bibinfo{volume}{18}, pp. \bibinfo{pages}{1--13},
  \doi{10.1017/S0956796807006326}.

\bibitemdeclare{article}{monads-moggi}
\bibitem{monads-moggi}
\bibinfo{author}{Eugenio Moggi} (\bibinfo{year}{1991}):
  \emph{\bibinfo{title}{{Notions of Computation and Monads}}}.
\newblock {\sl \bibinfo{journal}{Inf. Comput.}} \bibinfo{volume}{93}, pp.
  \bibinfo{pages}{55--92}, \doi{10.1016/0890-5401(91)90052-4}.

\bibitemdeclare{inproceedings}{comonads-ypnos}
\bibitem{comonads-ypnos}
\bibinfo{author}{Dominic Orchard}, \bibinfo{author}{Max Bolingbroke} \&
  \bibinfo{author}{Alan Mycroft} (\bibinfo{year}{2010}):
  \emph{\bibinfo{title}{Ypnos: {D}eclarative, {P}arallel {S}tructured {G}rid
  {P}rogramming}}.
\newblock In: {\sl \bibinfo{booktitle}{Proceedings of DAMP 2010}}, pp.
  \bibinfo{pages}{15--24}, \doi{10.1145/1708046.1708053}.

\bibitemdeclare{inproceedings}{types-constraints-unleashed}
\bibitem{types-constraints-unleashed}
\bibinfo{author}{Dominic~A. Orchard} \& \bibinfo{author}{Tom Schrijvers}
  (\bibinfo{year}{2010}): \emph{\bibinfo{title}{Haskell Type Constraints
  Unleashed}}.
\newblock In: {\sl \bibinfo{booktitle}{Proceedings of FLOPS 2010}}, pp.
  \bibinfo{pages}{56--71}, \doi{10.1007/978-3-642-12251-4\_6}.

\bibitemdeclare{inproceedings}{joinads-haskell11}
\bibitem{joinads-haskell11}
\bibinfo{author}{Tomas Petricek}, \bibinfo{author}{Alan Mycroft} \&
  \bibinfo{author}{Don Syme}: \emph{\bibinfo{title}{Extending {M}onads with
  {P}attern {M}atching}}.
\newblock In: {\sl \bibinfo{booktitle}{Proceedings of Haskell Symposium}},
  \bibinfo{series}{Haskell 2011}, \doi{10.1145/2096148.2034677}.

\bibitemdeclare{inproceedings}{joinads}
\bibitem{joinads}
\bibinfo{author}{Tomas Petricek} \& \bibinfo{author}{Don Syme}
  (\bibinfo{year}{2011}): \emph{\bibinfo{title}{Joinads: A Retargetable
  Control-Flow Construct for Reactive, Parallel and Concurrent Programming}}.
\newblock In: {\sl \bibinfo{booktitle}{Proceedings of PADL 2011}},
  \doi{10.1007/978-3-642-18378-2\_17}.

\bibitemdeclare{inproceedings}{formalism-lazy-sharing-semantics}
\bibitem{formalism-lazy-sharing-semantics}
\bibinfo{author}{S.~Purushothaman} \& \bibinfo{author}{Jill Seaman}
  (\bibinfo{year}{1992}): \emph{\bibinfo{title}{An Adequate Operational
  Semantics for Sharing in Lazy Evaluation}}.
\newblock In: {\sl \bibinfo{booktitle}{Proceedings of the 4th European
  Symposium on Programming}}, \doi{10.1007/3-540-55253-7\_26}.

\bibitemdeclare{}{monad-sttrans-package}
\bibitem{monad-sttrans-package}
\bibinfo{author}{Josef Svenningsson} (\bibinfo{year}{2011}):
  \emph{\bibinfo{title}{The STMonadTrans package}}.
\newblock
  \urlprefix\url{http://hackage.haskell.org/package/STMonadTrans-0.3.1}.

\bibitemdeclare{article}{comonads-dataflow-essence}
\bibitem{comonads-dataflow-essence}
\bibinfo{author}{Tarmo Uustalu} \& \bibinfo{author}{Varmo Vene}
  (\bibinfo{year}{2006}): \emph{\bibinfo{title}{The {E}ssence of {D}ataflow
  {P}rogramming}}.
\newblock {\sl \bibinfo{journal}{Lecture Notes in Computer Science}}
  \bibinfo{volume}{4164}, pp. \bibinfo{pages}{135--167},
  \doi{10.1007/11894100\_5}.

\bibitemdeclare{article}{comonads-notions}
\bibitem{comonads-notions}
\bibinfo{author}{Tarmo Uustalu} \& \bibinfo{author}{Varmo Vene}
  (\bibinfo{year}{2008}): \emph{\bibinfo{title}{Comonadic Notions of
  Computation}}.
\newblock {\sl \bibinfo{journal}{Electron. Notes Theor. Comput. Sci.}}
  \bibinfo{volume}{203}, pp. \bibinfo{pages}{263--284},
  \doi{10.1016/j.entcs.2008.05.029}.

\bibitemdeclare{inproceedings}{types-free-theorems-classes}
\bibitem{types-free-theorems-classes}
\bibinfo{author}{Janis Voigtl\"{a}nder} (\bibinfo{year}{2009}):
  \emph{\bibinfo{title}{{Free Theorems Involving Type Constructor Classes:
  Functional Pearl}}}.
\newblock In: {\sl \bibinfo{booktitle}{Proceedings of ICFP 2009}},
  \bibinfo{series}{ICFP 209}, \bibinfo{publisher}{ACM}, \bibinfo{address}{New
  York, NY, USA}, pp. \bibinfo{pages}{173--184}, \doi{10.1145/1596550.1596577}.

\bibitemdeclare{inproceedings}{monads-wadler}
\bibitem{monads-wadler}
\bibinfo{author}{Philip Wadler} (\bibinfo{year}{1990}):
  \emph{\bibinfo{title}{Comprehending Monads}}.
\newblock In: {\sl \bibinfo{booktitle}{Proceedings of Conference on LISP and
  Functional Programming}}, \bibinfo{series}{LFP 1990},
  \bibinfo{publisher}{ACM}, \bibinfo{address}{New York, NY, USA}, pp.
  \bibinfo{pages}{61--78}, \doi{10.1145/91556.91592}.

\bibitemdeclare{inproceedings}{formalism-cbn-dual-cbv}
\bibitem{formalism-cbn-dual-cbv}
\bibinfo{author}{Philip Wadler} (\bibinfo{year}{2003}):
  \emph{\bibinfo{title}{{Call-By-Value Is Dual to Call-By-Name}}}.
\newblock In: {\sl \bibinfo{booktitle}{Proceedings of ICFP}},
  \bibinfo{series}{ICFP 2003}, pp. \bibinfo{pages}{189--201},
  \doi{10.1145/944705.944723}.

\end{thebibliography}

\newpage
\appendix
\section{Equivalence proofs}
\label{sec:appendix-cbv-cbn-equivalence}

In this section, we show that our translation presented in Section~\ref{sec:abstracting-translation}
can be used to implement standard \emph{call-by-name} and \emph{call-by-value}. We 
prove that using an appropriate definition of \ensuremath{\Varid{malias}} from Section~\ref{sec:abstracting-aliasing}
gives the same semantics as the standard translations described by Wadler \cite{monads-wadler}.

\paragraph{Call-by-name.} The translation of types is the same for our translation 
and the \emph{call-by-name} translation. In addition, the rules for translating variable
access and lambda functions are also the same. This means that we only need to prove
that our translation of let-binding and function application are equivalent. When implementing
\emph{call-by-name} using our translation, we use the following definition of \ensuremath{\Varid{malias}}:

\begin{hscode}\SaveRestoreHook
\column{B}{@{}>{\hspre}l<{\hspost}@{}}%
\column{E}{@{}>{\hspre}l<{\hspost}@{}}%
\>[B]{}\Varid{malias}\;\Varid{m}\mathrel{=}\Varid{unit}\;\Varid{m}{}\<[E]%
\ColumnHook
\end{hscode}\resethooks
Using this definition and the \emph{left identity} monad law, we can now show that our 
\emph{call-by-alias} translation is equivalent to the translation from 
Figure~\ref{fig:translation-cbn}. The Figure~\ref{fig:equivalence-cbn} shows the equations
for function application and let-binding.

\paragraph{Call-by-value.} Proving that appropriate definition of \ensuremath{\Varid{malias}} gives a term
that corresponds to the one obtained using standard \emph{call-by-value} translation is
more difficult. In the \emph{call-by-value} translation, functions are translated to a
type $\tau_1 \rightarrow M~\tau_2$, while our translation produces functions of type
$M~\tau_1 \rightarrow M~\tau_2$. As a reminder, the definition of \ensuremath{\Varid{malias}} that gives the
\emph{call-by-value} behaviour is the following:

\begin{hscode}\SaveRestoreHook
\column{B}{@{}>{\hspre}l<{\hspost}@{}}%
\column{E}{@{}>{\hspre}l<{\hspost}@{}}%
\>[B]{}\Varid{malias}\;\Varid{m}\mathrel{=}\Varid{bind}\;\Varid{m}\;(\Varid{unit}\mathbin{\circ}\Varid{unit}){}\<[E]%
\ColumnHook
\end{hscode}\resethooks
To prove that the two translations are equivalent, we show that the following invariant holds:
when using the above definition of \ensuremath{\Varid{malias}} and our \emph{call-by-alias} translation, 
a monadic computation of type $M~\tau$ that is assigned to a variable $x$ always has
a structure $\ident{unit}~x_v$ where $x_v$ is a variable of type $\tau$. 

The sketch of the proof is shown in Figure~\ref{fig:equivalence-cbv}. We write $e_1 \cong e_2$
to mean that the expression $e_1$ in the \emph{call-by-alias} translation corresponds to an
expression $e_2$ in \emph{call-by-value} translation. This means that expressions of form $\ident{unit}~x_v$
translate to values $x$, variable declarations of $x_v$ (in lambda abstraction and let-binding) translate
to declarations of $x$ and all function values of type $M~\tau_1 \rightarrow M~\tau_2$ become
$\tau_1 \rightarrow M~\tau_2$.

\begin{figure}
\[
\begin{array}{lcl}
\transp{cba}{e_1 \: e_2}\\
\quad=~ \ident{bind}~\trans{cba}{e_1}~(\lambda f . \ident{bind}~(\ident{malias}~\trans{cba}{e_2})~f) ) 
  &&\textnormal{(translation)} \\ 
\quad=~ \ident{bind}~\trans{cba}{e_1}~(\lambda f . \ident{bind}~(\ident{unit}~\trans{cba}{e_2})~f) )
  &&\textnormal{(definition)}\\
\quad=~ \ident{bind}~\trans{cba}{e_1}~(\lambda f . f~\trans{cba}{e_2})
  &&\textnormal{(left identity)}\\
\quad=~ \ident{bind}~\trans{cbn}{e_1}~(\lambda f . f~\trans{cbn}{e_2})
  &&\textnormal{(induction hypothesis)}\\
\quad=~ \transp{cbn}{e_1 \: e_2}
  &\quad&\textnormal{(translation)}\\
\\[-0.5em]
\transp{cba}{\kvd{let} \: x = e_1 \: \kvd{in} \: e_2}\\
\quad=~ \ident{bind}~(\ident{malias}~\trans{cba}{e_1})~(\lambda x . \trans{cba}{e_2})
  &&\textnormal{(translation)} \\ 
\quad=~ \ident{bind}~(\ident{unit}~\trans{cba}{e_1})~(\lambda x . \trans{cba}{e_2})
  &&\textnormal{(definition)}\\
\quad=~ (\lambda x . \trans{cba}{e_2})~\trans{cba}{e_1}
  &&\textnormal{(left identity)}\\
\quad=~ (\lambda x . \trans{cbn}{e_2})~\trans{cbn}{e_1}
  &&\textnormal{(induction hypothesis)}\\
\quad=~ \transp{cbn}{\kvd{let} \: x = e_1 \: \kvd{in} \: e_2}
  &\quad&\textnormal{(translation)}
\end{array}
\]
\vspace{-1.5em}
\caption{Proving that using an appropriate \emph{malias} is equivalent to \emph{call-by-name}.}
\label{fig:equivalence-cbn}
\end{figure}

\begin{figure}
\[
\begin{array}{lcl}
\transp{cba}{x}\\
\quad=~ \ident{unit}~x_v
  &\quad&\textnormal{(assumption)}\\
\quad\cong~ {x}
  &\quad&\textnormal{(correspondence)}\\
\quad=~ \transp{cbv}{x}
  &\quad&\textnormal{(translation)}\\
\\
\transp{cba}{e_1 \: e_2}\\
\quad=~ \ident{bind}~\trans{cba}{e_1}~(\lambda f . 
    \ident{bind}~(\ident{malias}~\trans{cba}{e_2})~f) 
  &\quad&\textnormal{(translation)}\\
\quad=~ \ident{bind}~\trans{cba}{e_1}~(\lambda f . 
    \ident{bind}~(\ident{bind}~\trans{cba}{e_2}~(\ident{unit} \circ \ident{unit}))~f)
  &\quad&\textnormal{(definition)}\\
\quad=~ \ident{bind}~\trans{cba}{e_1}~(\lambda f . 
    \ident{bind}~\trans{cba}{e_2}~(\lambda x_v . \ident{bind}~(\ident{unit}~(\ident{unit}~x_v))~f))
  &\quad&\textnormal{(associativity)}\\
\quad=~ \ident{bind}~\trans{cba}{e_1}~(\lambda f . 
    \ident{bind}~\trans{cba}{e_2}~(\lambda x_v . f~(\ident{unit}~x_v)))
  &\quad&\textnormal{(left identity)}\\
\quad\cong~ \ident{bind}~\trans{cbv}{e_1}~(\lambda f . 
    \ident{bind}~\trans{cbv}{e_2}~(\lambda x . f~x)) 
  &\quad&\textnormal{(correspondence)}\\
\quad=~ \transp{cbv}{e_1 \: e_2}
  &\quad&\textnormal{(translation)}\\
\\
\transp{cba}{\kvd{let} \: x = e_1 \: \kvd{in} \: e_2}\\
\quad=~ \ident{bind}~(\ident{malias}~\trans{cba}{e_1})~(\lambda x . \trans{cba}{e_2})
  &&\textnormal{(translation)}\\
\quad=~ \ident{bind}~(\ident{bind}~\trans{cba}{e_1}~(\ident{unit} \circ \ident{unit}))~(\lambda x . \trans{cba}{e_2})
  &&\textnormal{(definition)}\\
\quad=~ \ident{bind}~\trans{cba}{e_1}~(\lambda x_v . \ident{bind}~(\ident{unit}~(\ident{unit}~x_v))~(\lambda x . \trans{cba}{e_2}))
  &&\textnormal{(associativity)}\\
\quad=~ \ident{bind}~\trans{cba}{e_1}~(\lambda x_v . (\lambda x . \trans{cba}{e_2})~(\ident{unit}~x_v))
  &&\textnormal{(left identity)}\\
\quad\cong~ \ident{bind}~\trans{cbv}{e_1}~(\lambda x . \trans{cbv}{e_2})
  &&\textnormal{(correspondence)}\\
\quad=~ \transp{cbv}{\kvd{let} \: x = e_1 \: \kvd{in} \: e_2}
  &\quad&\textnormal{(translation)}\\
\end{array}
\]
\caption{Proving that using an appropriate \emph{malias} is equivalent to \emph{call-by-value}.}
\label{fig:equivalence-cbv}
\end{figure}

\section{Proofs for two implementations}
\label{sec:appendix-cbv-cbn-proofs}

In this section, we prove that the two implementations of \ensuremath{\Varid{malias}} presented in 
Section~\ref{sec:abstracting-aliasing} obey the \ensuremath{\Varid{malias}} laws. We use the formulation
of monads based on a functor with additional operations \ensuremath{\Varid{join}} and \ensuremath{\Varid{unit}}. The 
proof relies on the following laws that hold about \ensuremath{\Varid{join}} and \ensuremath{\Varid{unit}}: 

\[
\begin{array}{rclcl}
\ident{map}~(g \circ f) &=& (\ident{map}~g) \circ (\ident{map}~f) &\quad&\textnormal{(functor)} \\
\ident{unit} \circ f  &=& \ident{map}~f \circ \ident{unit}    && \textnormal{(natural unit)}     \\
\ident{join} \circ \ident{map}~(\ident{map}~f) &=& \ident{map}~f \circ \ident{join}  && \textnormal{(natural join)}\\ 
\ident{join} \circ \ident{map}~\ident{join} &=& \ident{join} \circ \ident{join} && \textnormal{(assoc join)}\\ 
\ident{join} \circ \ident{unit} &=& \ident{join} \circ \ident{map}~\ident{unit} = \ident{id}  && \textnormal{(identity)} \\ 
\end{array}
\]
The first law follows from the fact that \ensuremath{\Varid{map}} corresponds to a functor. The next
two laws hold because \ensuremath{\Varid{unit}} and \ensuremath{\Varid{join}} are both natural transformations. Finally,
the last two laws are additional laws that are required to hold about monads (a precise 
definition can be found in Section~\ref{sec:theory}).

The proofs that the two definitions of \ensuremath{\Varid{malias}} (implementing \emph{call-by-name} and
\emph{call-by-value} strategies) are correct can be found in Figure~\ref{fig:proofs-cbn} and 
Figure~\ref{fig:proofs-cbv}, respectively. The proofs use the above monad laws. We use a
definition \ensuremath{\Varid{malias}\equiv \Varid{map}\;\Varid{unit}}, which is equivalent to the definition in Appendix~\ref{sec:appendix-cbv-cbn-equivalence}.
The figures include proofs for the four additional \ensuremath{\Varid{malias}} laws as defined in Section~\ref{sec:abstracting-malias}:
\emph{naturality}, \emph{associativity}, \emph{computationality} and \emph{identity}.

\begin{figure}
\[
\begin{array}{lcl}
\ident{map}~(\ident{map}~f) \circ \ident{malias}\\
\quad=~ \ident{map}~(\ident{map}~f) \circ \ident{unit}
  &\quad&\textnormal{(definition)}\\
\quad=~ \ident{unit} \circ \ident{map}~f
  &\quad&\textnormal{(naturality)}\\
\quad=~ \ident{malias} \circ \ident{map}~f
  &\quad&\textnormal{(definition)}\\
\\

\ident{map}~\ident{malias} \circ \ident{malias}\\
\quad=~ \ident{map}~\ident{unit} \circ \ident{unit}
  &\quad&\textnormal{(definition)}\\
\quad=~ \ident{unit}~\circ~\ident{unit}
  &\quad&\textnormal{(natural unit)}\\
\quad=~ \ident{malias}\circ\ident{malias}
  &\quad&\textnormal{(definition)}\\
\\

\ident{malias}\circ \ident{unit}\\
\quad=~ \ident{unit}\circ\ident{unit}
  &\quad&\textnormal{(definition)}\\
\\

\ident{join}\circ \ident{malias}\\
\quad=~ \ident{join}\circ\ident{unit}
  &\quad&\textnormal{(definition)}\\
\quad=~ \ident{id}
  &\quad&\textnormal{(identity)}\\
\end{array}
\]
\caption{Call-by-name definition of \ensuremath{\Varid{malias}} obeys the laws}
\label{fig:proofs-cbn}
\end{figure}

\begin{figure}

\[
\begin{array}{lcl}
\ident{map}~(\ident{map}~f) \circ \ident{malias}\\
\quad=~ \ident{map}~(\ident{map}~f) \circ \ident{map}~\ident{unit}
  &\quad&\textnormal{(definition)}\\
\quad=~ \ident{map}~(\ident{map}~f \circ \ident{unit})
  &\quad&\textnormal{(functor)}\\
\quad=~ \ident{map}~(\ident{unit} \circ f)
  &\quad&\textnormal{(natural unit)}\\
\quad=~ \ident{map}~\ident{unit} \circ \ident{map}~f
  &\quad&\textnormal{(functor)}\\
\quad=~ \ident{malias}~\circ~\ident{map}~f
  &\quad&\textnormal{(definition)}\\
\\

\ident{map}~\ident{malias} \circ \ident{malias}\\
\quad=~ \ident{map}~(\ident{map}~\ident{unit}) \circ (\ident{map}~\ident{unit})
  &\quad&\textnormal{(definition)}\\
\quad=~ \ident{map}~(\ident{map}~\ident{unit} \circ \ident{unit})
  &\quad&\textnormal{(functor)}\\
\quad=~ \ident{map}~(\ident{unit} \circ \ident{unit})
  &\quad&\textnormal{(natural unit)}\\
\quad=~ \ident{map}~\ident{unit} \circ \ident{map}~\ident{unit}
  &\quad&\textnormal{(functor)}\\
\quad=~ \ident{malias} \circ \ident{malias}
  &\quad&\textnormal{(definition)}\\
\\

\ident{malias}\circ \ident{unit}\\
\quad=~ \ident{map}~\ident{unit} \circ \ident{unit}
  &\quad&\textnormal{(definition)}\\
\quad=~ \ident{unit} \circ \ident{unit}
  &\quad&\textnormal{(natural unit)}\\
\\

\ident{join}\circ \ident{malias}\\
\quad=~ \ident{join} \circ \ident{map}~\ident{unit}
  &\quad&\textnormal{(definition)}\\
\quad=~ \ident{id}
  &\quad&\textnormal{(identity)}\\
\end{array}
\]

\caption{Call-by-value definition of \ensuremath{\Varid{malias}} obeys the laws}
\label{fig:proofs-cbv}
\end{figure}

\section{Typing preservation proof}
\label{sec:appendix-preserve-typing}

In this section, we show that our translation preserves typing. Given a well-typed term $e$
of type $\tau$, the translated term $\transp{cba}{e}$ is also well-typed and has a type
$\transp{cba}{\tau}$. In the rest of this section, we write $\transs{ - }$ for $\transp{cba}{ - }$. 
To show that the property holds, we use induction over the typing rules, using the 
fact that $\transs{\tau} = M~\tau'$ for some $\tau'$. The inductive construction of the 
typing derivation follows the following rules:

\begin{equation*}
\begin{array}{rcl}
\inference[]
  {}
  {\transs{\Gamma, x : \tau \vdash x : \tau}}
&=&
\inference
  {}
  {\transs{\Gamma}, x : M~\transs{\tau} \vdash x : M~\transs{\tau}}
\\
\\
\inference[]
  {\transs{\Gamma, x : \tau_1 \vdash e : \tau_2}}
  {\transs{\Gamma \vdash \lambda x . e : \tau_1 \rightarrow \tau_2}}
&=&  
\inference
  {\transs{\Gamma}, x : M~\transs{\tau_1} \vdash \transs{e} : M~\transs{\tau_2}}
  {\transs{\Gamma} \vdash \ident{unit}~(\lambda x . \transs{e}) : M~(M~\transs{\tau_1} \rightarrow M~\transs{\tau_2})}
\\
\\
\inference[]
  {\transs{\Gamma \vdash e_1 : \tau_1 \rightarrow \tau_2} \quad \transs{\Gamma \vdash e_2 : \tau_1}}
  {\transs{\Gamma \vdash e_1~e_2 : \tau_2 }}
&=&  
\inference
  {\transs{\Gamma} \vdash \transs{e_1} : M~(M~\transs{\tau_1} \rightarrow M~\transs{\tau_2}) \quad \Gamma \vdash \transs{e_2} : M~\transs{\tau_1}}
  {\Gamma \vdash \ident{bind}~\transs{e_1}~(\lambda f . \ident{bind}~(\ident{malias}~\transs{e_2})~f) ) : M~\transs{\tau_2} }
\\
\\
\inference[]
  {\transs{\Gamma \vdash e_1 : \tau_1} \quad \transs{\Gamma, x : \tau_1 \vdash e_2 : \tau_2}}
  {\transs{\Gamma \vdash \kvd{let}~x = e_1~\kvd{in}~e_2 : \tau_2 }}
&=&  
\inference
  {\transs{\Gamma} \vdash \transs{e_1} : M~\transs{\tau_1} \quad \transs{\Gamma}, x : M~\transs{\tau_1} \vdash \transs{e_2} : M~\transs{\tau_2}}
  {\transs{\Gamma} \vdash \ident{bind}~(\ident{malias}~\transs{e_1})~(\lambda x . \transs{e_2}) : M~\transs{\tau_2} }
\end{array}
\end{equation*}

\end{document}